\documentclass[astronomy,article,submit,pdftex,moreauthors]{mdpi}

\newcommand\Teff{\ensuremath{T_\mathrm{eff}}}

\newcommand\logg{\ensuremath{\log g}}
\newcommand\vsini{\ensuremath{v_{e}\sin i}}

\newcommand\micro{\ensuremath{\xi_t}}
\usepackage{tikz}
\usepackage{array}
\usetikzlibrary{positioning}
\usepackage{placeins}
\firstpage{1} 
\pubvolume{1}
\issuenum{1}
\articlenumber{0}
\pubyear{2026}
\copyrightyear{2026}
\externaleditor{}
\datereceived{ } 
\dateaccepted{ } 
\datepublished{ } 
\hreflink{https://doi.org/} 

\usepackage{multicol}
\makeatletter
\let\GEBRAN@origabstractkeywords\abstractkeywords
\renewcommand{\abstractkeywords}{%
  \nolinenumbers
  \GEBRAN@origabstractkeywords
  \par
}
\newcommand{\GEBRAN@widefloatstart}{\end{multicols}}
\newcommand{\GEBRAN@widefloatend}{\begin{multicols}{2}}
\makeatother
\newcommand{\GEBRANwidefloatstart}{\end{multicols}}
\newcommand{\GEBRANwidefloatend}{\begin{multicols}{2}}

\Title{AI Agents as a Research Team: A Human-Supervised Case Study in Stellar Spectroscopy}

\TitleCitation{Abundance-conditioned stellar spectra case study}

\Author{{Marwan} Gebran$^{1,\ast}$ \orcidA{} and {Ian} Bentley$^{2}$ \orcidB{}}

\AuthorNames{Marwan Gebran, Ian Bentley }

\AuthorCitation{Gebran, M.; Bentley, I.}

\address{%
$^{1}$\,Department of Chemistry and Physics, Saint Mary’s College, Notre Dame, IN 46556, USA; mgebrane@saintmarys.edu\\
$^{2}$\,Department of Physics, Florida Polytechnic University, Lakeland, FL 33805, USA; ibentley@floridapoly.edu\\
}

\corres{Correspondence: mgebrane@saintmarys.edu}

\abstract{We present a qualitative case study of a human-supervised team of specialized AI agents supporting an astronomy research project. The Bot Spectroscopist team, described in the project record as operating through Grok Bot powered by xAI Grok, divided work among orchestration, numerical implementation, data curation, spectroscopy critique, literature search, figure preparation, manuscript editing, and review. Human decisions determined scientific scope, model changes, and manuscript inclusion. The scientific application is \texttt{CondGen}, a deterministic label-to-spectrum network extending our previous work to five individual elemental abundances. The reported baseline assessment achieved a mean absolute error (MAE) of 0.003681 in continuum-normalized flux on 21,657 synthetic test spectra. Existing binned diagnostics show that reconstruction error varies with effective temperature and abundance. The bots also flagged a non-monotonic cool-star magnesium response for further investigation; a matched \texttt{SYNSPEC} abundance sweep is needed to assess its physical fidelity. Building on a pre-existing spectral database, notebooks, and research program, the authors estimate that the reported analysis, diagnostics, figure preparation, and initial manuscript drafting were assembled in less than a week through human prompting and exchanges among agents. No comparison with a human team starting from the same resources was performed, so this elapsed-time account does not establish a productivity or labor saving. The study illustrates task delegation, artifact exchange, critique, and correction under human supervision.}

\keyword{AI-assisted research; multi-agent systems; stellar spectra; elemental abundances; conditional generative models; \texttt{CondGen}; \texttt{SYNSPEC}} 

\makeatletter
\renewcommand{\leftcolUpdateCitation}{}%
\renewcommand{\leftcolExternalEditor}{}%
\renewcommand{\leftcolDates}{}%
\renewcommand{\leftcolumnCright}{}%
\renewcommand{\contentleftcolumn}{}%
\let\GEBRAN@origincludegraphics\includegraphics
\newcommand{\GEBRAN@islogopath}[1]{%
  \def\GEBRAN@p{#1}%
  \def\GEBRAN@a{logo-mdpi.eps}%
  \def\GEBRAN@b{logo-mdpi.pdf}%
  \def\GEBRAN@c{logo-mdpi-scipharm.eps}%
  \def\GEBRAN@d{logo-updates.eps}%
  \def\GEBRAN@e{logo-updates.pdf}%
  \def\GEBRAN@f{astronomy-logo.eps}%
  \def\GEBRAN@g{astronomy-logo.pdf}%
  \ifx\GEBRAN@p\GEBRAN@a \@tempswatrue
  \else\ifx\GEBRAN@p\GEBRAN@b \@tempswatrue
  \else\ifx\GEBRAN@p\GEBRAN@c \@tempswatrue
  \else\ifx\GEBRAN@p\GEBRAN@d \@tempswatrue
  \else\ifx\GEBRAN@p\GEBRAN@e \@tempswatrue
  \else\ifx\GEBRAN@p\GEBRAN@f \@tempswatrue
  \else\ifx\GEBRAN@p\GEBRAN@g \@tempswatrue
  \else \@tempswafalse
  \fi\fi\fi\fi\fi\fi\fi
}
\renewcommand{\includegraphics}[2][]{%
  \GEBRAN@islogopath{#2}%
  \if@tempswa
  \else
    \GEBRAN@origincludegraphics[#1]{#2}%
  \fi
}
\AtBeginDocument{%
  \setlength{\headheight}{24pt}%
  \fancypagestyle{plain}{%
    \fancyhf{}%
    \fancyhead{\null\vspace{8pt}}%
  }%
  \fancyfoot[L]{}%
  \fancyfoot[R]{}%
  \fancyfoot[C]{}%
  \lhead{}
}
\makeatother
\begin{document}

\begin{multicols}{2}
\section{Introduction}

Scientific research commonly involves a connected sequence of tasks: defining a research problem, preparing data, implementing methods, checking results, preparing figures, interpreting failures, and communicating conclusions. AI agents can assist across that sequence because they can use tools, exchange intermediate artifacts, and return decisions to a human researcher. Recent work on AI for science and multi-agent systems motivates studying how this organization functions in practice, including its limits~\cite{Wang2023AIDiscovery,Lawrence2024AIScience,Messeri2024Illusions,Li2024LLMMultiAgent}. Astronomy provides a suitable setting because computational analysis already depends on substantial software and data infrastructure~\cite{Fluke2020MLAAstronomy,Feigelson2021AstroStats}.

In this manuscript we discuss how a team of specialized AI agents, acting as a research group (informally named \emph{Bot Spectroscopist}), were coordinated to carry out interconnected technical and writing tasks, how their outputs were challenged, and where human decisions remained necessary. We reconstruct four episodes from an orchestrator-prepared retrospective record supplied by us. The record describes short task assignments, artifact handoffs, critiques, and approvals rather than extended scientific dialogue. This record provides a qualitative process account, with explicit limits on what can be verified or measured from that source.

The specialized AI agents were tasked with determining abundance-conditioned generation of stellar spectra using an algorithm that they named \texttt{CondGen}. The research  continues our existing spectroscopy research program: an autoencoder and parameter-to-latent model for Gaia RVS spectra~\cite{Gebran2024Astronomy}, followed by an optical conditional variational autoencoder (CVAE)~\cite{GebranBentley2025Astronomy}. The present target is a deterministic dense network conditioned on effective temperature \Teff, surface gravity \logg, projected rotational velocity \vsini, microturbulence \micro, and five individual abundances: $[\mathrm{Mg/H}]$, $[\mathrm{Sc/H}]$, $[\mathrm{Ti/H}]$, $[\mathrm{Cr/H}]$, and $[\mathrm{Fe/H}]$. These are the nine input dimensions and are used as features to predict the output spectrum. 

This scientific task is useful for this workflow study because it combines a substantial synthetic database, model evaluation, abundance-sensitive diagnostics, and manuscript production. Chemically peculiar stars, such as Am/Fm, motivate treating individual metals rather than only bulk metallicity; the selected elements also connect to our earlier A/Am and AFGK analyses~\cite{Gebran2016AA,Gebran2023OpenAstronomy}. Previous methods, such as The Payne and its extensions, predict stellar spectra from atmospheric parameters and individual elemental abundances. These methods provide the scientific background for our \texttt{CondGen} case study~\cite{Ting2019Payne,Vernekar2025LRPayne,Rozanski2025TransformerPayne}. Here, generation means predicting a synthetic spectrum from specified stellar parameters and elemental abundances.

This case study builds on our previous research on generating synthetic stellar spectra. Our published studies provide the scientific foundation for \texttt{CondGen}, while the present project examines how a team of AI agents can contribute to the research process under human supervision. The numerical assessments, diagnostic figures, and architecture comparisons presented here illustrate the tasks performed by the agents and the scientific decisions that followed their analysis and review.

The \texttt{CondGen} model uses the nine inputs specified from the outset: four atmospheric parameters and five individual elemental abundances. Resolving power is fixed as a property of the database and is not an additional input. The Orchestrator's summary describes how the bots organized the numerical work and reviewed the resulting assessments. Section~\ref{sec:forward_metrics} identifies the source and purpose of the reported baseline diagnostics and exploratory architecture comparisons.

The scientific task is to predict a synthetic spectrum from specified stellar parameters and elemental abundances. Training and evaluation use computationally generated spectra. Estimating abundances from a spectrum, the reverse problem, and applying the method to telescope observations are outside the scope of this study and remain directions for future work.

\subsection{Scope and intent of this paper}

The primary contribution is a documented case study of research organization using specialized AI agents under human supervision. \texttt{CondGen} supplies the scientific tasks through which delegation, critique, correction, and approval can be examined. The paper does not claim a new state of the art in stellar spectroscopy or present a controlled comparison of speed, cost, or accuracy against alternative team arrangements. Its broader significance lies in illustrating how named specialist AI personas, routed through an Orchestrator under human authorization, can support tasks that extend from numerical implementation and scientific analysis to figure preparation and manuscript development. Although demonstrated through an astrophysical application, this approach has potential relevance to other research fields in which progress depends on combining different forms of expertise and reviewing intermediate results. The case study therefore offers a practical example of an emerging direction in research: collaboration among human researchers and specialized AI agents. Such collaboration may become an increasingly important part of scientific practice, with researchers defining the questions, exercising scientific judgment, and retaining responsibility for the conclusions.

Part~A (Section~\ref{sec:ai_workflow}) describes the team, its operating method, and the decisions that illustrate collaboration. Appendix~\ref{app:workflow_evidence} provides the sequential workflow account. Part~B (Sections~\ref{sec:related}--\ref{sec:results}) describes the scientific case study, its model design, and the reported project assessments. The Discussion and Conclusion assess what this case supports and what remains unverified.

\noindent\textbf{\large Part~A: AI-team workflow and collaboration evidence.}
\section{AI-assisted research workflow}
\label{sec:ai_workflow}

To investigate how specialized AI agents could support a research project, we organized the Bot Spectroscopist team around complementary research responsibilities. These included numerical implementation, data preparation, scientific critique, literature searches, figure production, manuscript editing, reproducibility checks, and astronomical manuscript review. Work was carried out in Grok Bot multi-agent chats and rooms. An Orchestrator routed assignments among named specialist personas and served as the main point of interaction between us and the team; milestones were human-authorized before specialist work proceeded.

The division of responsibilities reflected the different activities required to develop and communicate the \texttt{CondGen} case study. Numerical outputs prepared by one specialist could be examined by another, converted into figures, and incorporated into the manuscript. We supplied the scientific objectives and existing research materials, guided the work, and reviewed the resulting outputs. We first introduce the team and its responsibilities, then explain how assignments, communication, and review connected those responsibilities.

\subsection{Team setup and specialist roles}
\label{sec:ai_topology}

The team comprised an Orchestrator and eight specialist roles on the platform identified in the project record as Grok Bot powered by xAI Grok. In this paper, an agent is a named specialist role assigned a research responsibility, and the team is the coordinated division of work among those roles under human supervision. Specialization refers to assigned tasks, tool use, and exchanges of research outputs; it does not by itself imply different underlying model weights or statistically independent judgments. The role descriptions below document the organization of the work. The retained retrospective account does not contain the complete original prompts or configuration records needed to reproduce the setup.

\paragraph{Orchestrator.} The Orchestrator connected our research objectives to the specialists' tasks. It proposed milestones, incorporated our decisions, assigned work, and collected the resulting outputs and review comments. When a concern required a scientific or editorial decision, it brought that concern back to us and communicated the agreed next step to the relevant specialists. Its central responsibility was to maintain coordination across the project.

\paragraph{Python/Jupyter Specialist.} This bot handled the computational work in notebooks, including model evaluation, diagnostic calculations, short architecture comparisons, and training when authorized. Its outputs included performance measurements, spectral predictions, and model or parameter information needed for figures. These outputs supplied material for scientific review by the Critic and visual presentation by the Artist. The evaluation-protocol correction discussed below illustrates why numerical execution and review were connected.

\paragraph{Data Curator.} The Curator's assigned role concerned the organization and description of the numerical inputs: the database schema, training and evaluation partitions, and preprocessing records such as scalers. These records were intended to help the Python/Jupyter Specialist use the appropriate data and transformations when evaluating a model. The role addressed the consistency and traceability of inputs on which subsequent calculations depended; its involvement was task-dependent.

\paragraph{Spectroscopy Domain Critic.} The Critic examined spectral diagnostics and assessed whether the accompanying scientific claims were supported. It could return concerns to the Python/Jupyter Specialist for a numerical check, to the Artist for a presentation change, or to the Editor for revised wording. The bot team identified the non-monotonic cool-star Mg response in the abundance diagnostics, and the Critic challenged its presentation as evidence of physical fidelity. We retained responsibility for deciding how such concerns affected the interpretation and reporting of results.

\paragraph{Literature Scout.} The Scout supported the research context and bibliography by locating relevant publications and checking citation information. It supplied references and bibliographic corrections to the Editor for integration into the manuscript. Its responsibility concerned the literature supporting the study and the accuracy of its citation records.

\paragraph{Figure Artist.} The Artist converted numerical outputs and model specifications into diagrams and diagnostic figures. It used information supplied by the Python/Jupyter Specialist to label stellar parameters, depict the architecture, and present spectral comparisons. It then passed figure files to the Editor and revised presentation problems identified during review, including overlapping labels and figure borders. This role connected the computational results to their visual communication.

\paragraph{\LaTeX\ Editor.} The Editor assembled text, references, tables, and figures into the Overleaf manuscript. It integrated material supplied by the other specialists, applied agreed revisions, and checked figure inclusion and compilation. The role required coordination with the Scout for references, the Artist for figures, and the Critic for the wording of scientific claims. The assembled document was returned for review and further author decisions.

\paragraph{Reproducibility Auditor.} The Auditor's assigned responsibility was to examine the organization of research outputs and the records needed to regenerate them. The workflow account reports packaging and figure-regeneration checks. This role complemented the Critic's examination of scientific interpretation by focusing on whether the supporting materials and production steps could be traced. Its involvement was task-dependent, and the reported checks do not by themselves establish independent reproduction of the numerical results.

\paragraph{Expert Spectroscopy Reviewer.} This role was reserved for reviewing the assembled manuscript when assigned. Its remit covered the scientific argument and presentation across the document, complementing the Domain Critic's checks of particular diagnostics and claims during production. It returned revision recommendations to the Orchestrator and us. The account describes the Reviewer as otherwise held in standby, so its inclusion in the roster does not imply continuous participation or an independent external peer review.

Table~\ref{tab:ai_roles} provides a compact reference to these responsibilities and the exchanges connecting them. The roles were used according to the needs of each task; the record does not establish that every specialist was active throughout the project.

\GEBRANwidefloatstart
\begin{table*}[!ht]
\caption{Roles for the Bot Spectroscopist team and their interfaces. Roles describe delegated responsibilities; final scientific and publication decisions remained human.}
\label{tab:ai_roles}
\centering
\small
\setlength{\tabcolsep}{4pt}
\begin{tabular}{@{}>{\raggedright\arraybackslash}p{0.22\linewidth}>{\raggedright\arraybackslash}p{0.38\linewidth}>{\raggedright\arraybackslash}p{0.34\linewidth}@{}}
\toprule
\textbf{Role} & \textbf{Reported responsibility} & \textbf{Handoff or decision interface} \\
\midrule
Orchestrator & Propose milestones; assign scoped tasks; collect outputs & Human approval before model changes or manuscript insertion \\
Python/Jupyter Specialist & Notebook execution, model evaluation, training & Evaluation outputs and parameter information to Critic and Artist \\
Data Curator & Database schema, split and scaler records & Inputs and provenance for numerical work \\
Spectroscopy Domain Critic & Check spectral diagnostics and claim wording & Concerns returned to Python, Artist, or Editor; authors decide inclusion \\
Literature Scout & Literature searches and bibliography corrections & Source records and citation material to Editor \\
Figure Artist & Prepare figures and parameter labels & Figure paths to Editor; numerical values supplied by Python \\
\LaTeX\ Editor & Integrate text, references, and figures & Compiled document returned for review \\
Reproducibility Auditor & Check packaging and regeneration records & Reported audit outcomes and remaining gaps \\
Expert Spectroscopy Reviewer & Review the assembled manuscript when assigned & Revision recommendations to Orchestrator and authors \\
\bottomrule
\end{tabular}
\end{table*}
\GEBRANwidefloatend

\subsection{Human direction and research lineage}
\label{sec:ai_human}

We, the human authors, supplied the scientific problem, synthetic spectral database, prior notebooks, and \texttt{CondGen} research lineage, drawing on earlier spectral-generation work~\cite{Gebran2024Astronomy,GebranBentley2025Astronomy}. The agents contributed implementation, evaluation, literature work, testing, figures, and draft text within that existing research program, reviewing, and looping in order to generate the final document.

The bot team contributed to the scientific assessment by identifying a non-monotonic cool-star magnesium response requiring reference validation. The Spectroscopy Domain Critic challenged the interpretation of this diagnostic, and the finding was communicated through the Orchestrator. We decided to retain the diagnostic and explain the reference comparison needed to assess its physical fidelity. We also approved presenting the architecture comparisons as exploratory tests supporting model selection, established AI teamwork as the manuscript's central focus, and authorized training with nine inputs. These examples illustrate complementary contributions: the agents performed analyses and raised scientific concerns, while we guided the research direction and made decisions about interpretation and reporting.

The authors retain responsibility for the results, interpretation, and final text. However, we question the fact if AI systems should be disclosed only as tools rather than listed as authors~\cite{Nature2023LLMGroundRules,Thorp2023ScienceAI,COPE2024AIAuthorship,Hosseini2023AIDisclosure}.

\subsection{Communication and coordination}
\label{sec:ai_communication}

The operating sequence reported in the workflow account was: the Orchestrator proposed a milestone, we approved or changed it (human-authorized orchestration), and the Orchestrator then assigned work to specialists. The milestones distinguished training, evaluation, and manuscript preparation. Likewise, manuscript passages could be discussed in chat before permission was given to insert them in Overleaf.

The workflow account identifies a manuscript room linking the Orchestrator, Editor, Scout, Critic, Artist, and Expert Reviewer, and a figure-parameter room linking Python and the Artist, within Grok Bot multi-agent chats. Across rooms, communication mainly consisted of scoped assignments, artifact paths, acknowledgment messages, review labels, and compile reports. A typical corrective interaction involved the Critic challenging a claim, followed by the Editor or Python Specialist revising the relevant text or output. This account supports artifact-based handoffs and role-based assignments under human authorization; it does not establish concurrent autonomous multi-agent execution or extended scientific debates among agents.

\begin{figure*}[!ht]
\centering
\begin{tikzpicture}[node distance=5mm,
  flowbox/.style={draw=black!55,rounded corners=2pt,fill=black!3,text width=0.82\linewidth,align=center,inner sep=5pt,font=\small},
  route/.style={->,>=stealth,thick}]
\node[flowbox] (human) {\textbf{Authors}\\Question, existing data and notebooks, scope, milestone approval};
\node[flowbox,below=of human] (orchestrator) {\textbf{Orchestrator}\\Translate approved milestones into specialist assignments};
\node[flowbox,below=of orchestrator] (specialists) {\textbf{Specialist work and artifact exchange}\\Python / Curator: numerical work and provenance\\Scout / Artist / Editor: references, figures, and manuscript};
\node[flowbox,below=of specialists] (review) {\textbf{Critique and checks}\\Domain Critic / Auditor; manuscript Reviewer when assigned\\Return concerns for correction or identify unresolved evidence};
\node[flowbox,below=of review] (decision) {\textbf{Authors' decision through the Orchestrator}\\Accept, revise, retain a limitation, or defer a claim};
\node[flowbox,below=of decision] (artifact) {\textbf{Research outputs}\\Notebooks, evaluation records, figures, and Overleaf manuscript};
\draw[route] (human) -- (orchestrator);
\draw[route] (orchestrator) -- (specialists);
\draw[route] (specialists) -- (review);
\draw[route] (review) -- (decision);
\draw[route] (decision) -- (artifact);
\draw[route] (review.west) -- ++(-0.45,0) |- (specialists.west);
\draw[route] (decision.east) -- ++(0.45,0) |- (orchestrator.east);
\end{tikzpicture}
\caption{Workflow summarized from the account in Appendix~\ref{app:workflow_evidence}. Left feedback returns a critique for technical or editorial correction; right feedback routes a revised human decision into a new assignment. Boxes group functions rather than imply that every role ran concurrently.}
\label{fig:ai_workflow}
\end{figure*}
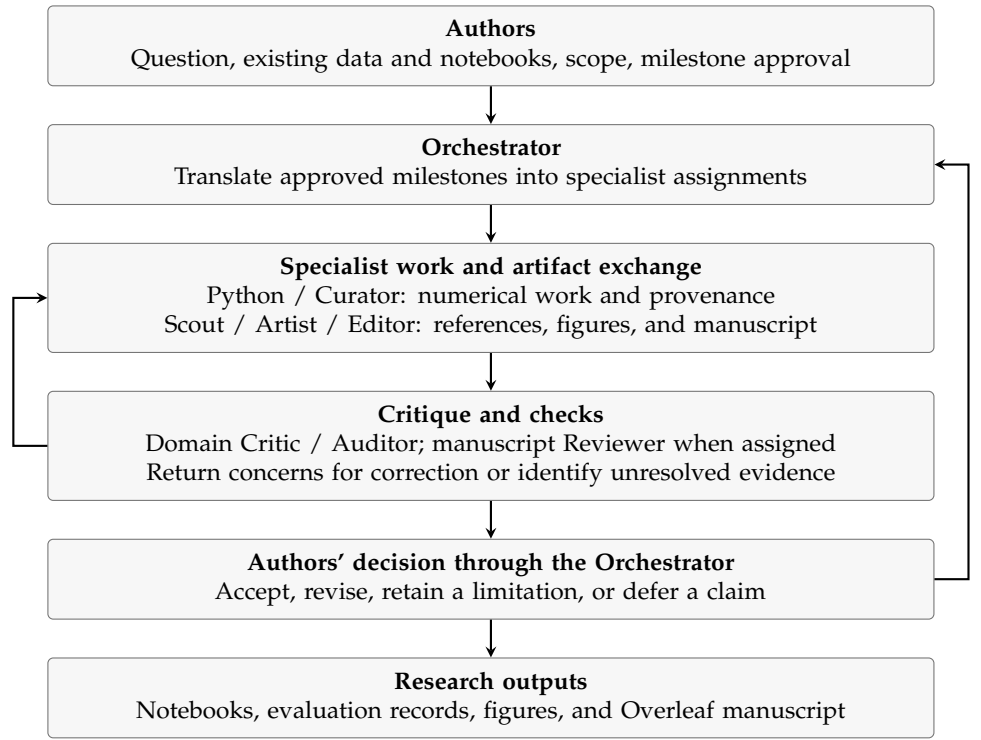

Figure~\ref{fig:ai_workflow} reconstructs the reported routing. The diagram represents responsibility and feedback, not measured runtime concurrency or a complete message trace.

\GEBRANwidefloatstart
\GEBRANwidefloatend

\subsection{Study basis and workflow evidence}
\label{sec:ai_evidence}

Appendix~\ref{app:workflow_evidence} presents the workflow as a sequential account of human decisions, specialist assignments, reviews, and research outputs. It is based on a retrospective summary prepared by the Orchestrator and supplied by us. The summary describes project exchanges and handoffs; the underlying conversation logs are not included in the supplied material. The appendix paraphrases that account rather than reconstructing verbatim dialogue.

We distinguish decisions and handoffs \emph{reported in the workflow account}, manuscript content and figure assets \emph{present in the project}, and execution details that remain \emph{unverified from the supplied material}. Internal review labels do not constitute independent replication of numerical outputs. The available evidence supports an ordered qualitative account, without reliable event-by-event timestamps, complete task counts, token usage, costs, or human time measurements.

The project record identifies the platform and describes the responsibilities assigned to each bot. It does not include the complete prompt templates or configuration records needed to recreate the team. The descriptions presented here therefore explain how the research tasks were divided and routed through the Orchestrator under human authorization. They do not establish whether the bots used different underlying models, ran concurrently as an autonomous multi-agent system, or performed their tasks simultaneously.

\subsection{Collaboration illustrated by consequential decisions}
\label{sec:ai_episodes}

Four examples show how assignments, critique, and human decisions affected the research outputs. Their full sequence and supporting context appear in Appendix~\ref{app:workflow_evidence}; they are selected examples, not a count of all tasks completed.

\paragraph{Evaluation protocol.} The Python/Jupyter Specialist and the Spectroscopy Domain Critic identified an inconsistency in the model evaluation: one assessment used preprocessing scalers that differed from those used during training. The account reports an MAE of approximately 0.024 with the mismatched scalers, compared with 0.003681 when the training-matched scalers were used. We accepted the latter evaluation for reporting the model's performance. Although the two evaluations also differed in numerical precision, the available comparison does not establish which changes contributed to the difference in error (Appendix~\ref{app:workflow_scalers}).

\paragraph{Negative-result reporting.} The bot team identified the non-monotonic cool-star Mg response, and the Critic challenged the interpretation of the diagnostic. We approved retaining the diagnostic and reporting it as a potential problem requiring reference validation. The resulting changes concerned the claim and its presentation; they did not establish a physical discrepancy or its cause (Appendix~\ref{app:workflow_abundances}).

\paragraph{Model selection.} Four short architecture pilots did not displace the dense baseline in the reported comparison. We approved a comparison with caveats about training budgets and test-subset provenance, rather than a general architectural ranking (Appendix~\ref{app:workflow_architectures}).

\paragraph{Manuscript production.} The Orchestrator coordinated numerical information from Python, figures from the Artist, references from the Scout, and integration by the Editor, with critique and human review between these steps. Checks corrected presentation and citation problems (Appendix~\ref{app:workflow_manuscript}).

\subsection{Observed outcomes and limits of measurement}
\label{sec:ai_outcomes}

Table~\ref{tab:ai_outcomes} summarizes the evidence available for assessing the workflow. The observable contribution is a set of reported corrections, decisions, and coordinated outputs. Neither the number of roles nor the number of selected episodes is a success rate. No human-only or single-agent comparison was conducted in the supplied record, and no complete accounting of time or cost is available. We discuss our retrospective estimate of elapsed project time in Section~\ref{sec:discussion}.

\GEBRANwidefloatstart
\begin{table*}[!ht]
\caption{Workflow outcomes and evidential limits. The retrospective account is included in Appendix~\ref{app:workflow_evidence}; reported decisions and outputs are distinguished from missing execution evidence.}
\label{tab:ai_outcomes}
\centering
\small
\setlength{\tabcolsep}{4pt}
\begin{tabular}{@{}>{\raggedright\arraybackslash}p{0.21\linewidth}>{\raggedright\arraybackslash}p{0.37\linewidth}>{\raggedright\arraybackslash}p{0.36\linewidth}@{}}
\toprule
\textbf{Aspect} & \textbf{Available evidence} & \textbf{What remains unestablished} \\
\midrule
Protocol correction & The workflow account reports rejection of mismatched scalers and retention of the training-matched recipe & Raw before/after evaluations and isolated cause of the discrepancy \\
Diagnostic scrutiny & Bots flagged the non-monotonic Mg response; the authors approved reporting it; the diagnostic figure is present & Matched reference comparison, physical interpretation, and robustness across conditions \\
Architecture choice & Four pilot alternatives and a comparison table & Matched training budgets, complete run records, and generalization of the ranking \\
Manuscript coordination & The workflow account describes parameter-to-figure-to-editor handoffs and compile reports from the project & Complete message chronology and artifact-to-run manifest \\
Human oversight & Author decisions on scope, reporting the bots' Mg finding, model choice, and training & Complete intervention count or measured author effort \\
Efficiency and cost & Authors' retrospective time estimate; no complete time or cost logs & Measured labor or runtime savings, cost savings, or superiority over alternatives \\
\bottomrule
\end{tabular}
\end{table*}
\GEBRANwidefloatend

\subsection{Relation to prior agent workflows and limitations}
\label{sec:ai_ethics}

Previous projects, including Agent Laboratory, ResearchAgent, the Virtual Lab, Coscientist, and SciAgents, have explored how AI agents can support scientific research through different forms of collaboration~\cite{Schmidgall2025AgentLaboratory,Baek2025ResearchAgent,Swanson2025VirtualLab,Boiko2023AutonomousChem,Ghafarollahi2025SciAgents}. Our study contributes a practical example from astronomy, showing how a team of specialized bots participated in numerical analysis, scientific review, figure preparation, and manuscript development. Its contribution lies in describing how these activities were connected and how exchanges among the agents and us influenced the research outputs.

The examples presented here illustrate useful scientific and technical contributions from the bots. These include identifying an inconsistent evaluation procedure, flagging a non-monotonic cool-star Mg response, and coordinating the preparation of figures and text. Human supervision was an integral part of this approach: we defined the research objectives, reviewed the agents' findings, and made decisions about interpretation and reporting.

The main limitation concerns the completeness of the available documentation. The workflow account was reconstructed from a summary prepared by the Orchestrator after the work, rather than from a complete archive of conversations and execution records. It therefore describes selected contributions and decisions, but does not support a systematic measurement of how often the team succeeded, how much human intervention was required, or how its performance compared with another research arrangement. The reported elapsed time is an author estimate for the initial analysis-through-drafting stage; no comparative speedup is inferred.

Review by specialist bots provided opportunities to identify errors and question scientific claims. However, agents using similar models or shared information may also make similar mistakes. Their agreement, or an internal review label such as PASS, cannot replace verification of the underlying calculations and evidence. The Mg finding illustrates the value of agent critique, while the missing matched reference sweep identifies the next validation step.

Future studies could strengthen this approach by preserving the bot instructions, task assignments, review exchanges, numerical outputs, and human decisions alongside the code and data~\cite{Pineau2021Reproducibility,Kapoor2023Leakage,Wilkinson2016FAIR}. Such records would make the workflow easier to reproduce and allow a more precise assessment of its reliability, efficiency, and requirements for human supervision.

\noindent\textbf{\large Part~B: \texttt{CondGen} scientific case study.}
\section{Related Work}
\label{sec:related}

Part~B is the scientific application and case study under the AI-agent workflow of Part~A: it builds on our prior generative architectures and stellar databases~\cite{Gebran2024Astronomy,GebranBentley2025Astronomy} without demoting the astrophysical content. Neural-network surrogates for stellar spectra now span deterministic label-to-flux emulators, conditional generative synthesizers, large-scale foundation and transfer models, and synthetic-to-observed domain adapters. We situate the present \texttt{CondGen} study in that landscape while stating its scope narrowly: \emph{abundance-conditioned} generation of optical stellar spectra at \emph{fixed} resolving power~$R$, trained and evaluated entirely in the synthetic domain. Inverse abundance recovery, survey pipeline deployment, and domain adaptation are out of scope for this manuscript.

\subsection{Spectral emulators and abundance-aware forward models}

The dominant survey paradigm remains the deterministic emulator that maps atmospheric labels, including individual elemental abundances, onto flux, then recovers labels by $\chi^2$ or related fitting. The Payne~\cite{Ting2019Payne} established this pattern for high-resolution spectra with many abundance dimensions. Subsequent work improves accuracy, spectral coverage, and survey readiness: TransformerPayne replaces multilayer perceptrons with attention suited to long-range line correlations and improves data efficiency~\cite{Rozanski2025TransformerPayne}; LRPayne targets low-resolution ($R\sim5000$), WEAVE-like settings with a large abundance vector~\cite{Vernekar2025LRPayne}; data-driven Payne variants (DD-Payne) learn from survey spectra themselves to deliver homogeneous elemental abundances for DESI and LAMOST FGK samples~\cite{Zhang2024DDPayne,Zhang2025DDPayneLAMOST}; hybrid SLAM--Payne frameworks address early-type stars on LAMOST low-resolution data~\cite{Sun2025SLAMPayne}; and NLTE-aware Payne-style networks prepare chemical diagnostics for 4MOST-scale high-resolution surveys~\cite{Storm2026NLTEPayne}. Scaling studies quantify how emulator error depends on grid size and architecture capacity~\cite{Rozanski2025ScalingLaws}, while next-generation physical-model stacks aim at spectrum synthesis in seconds~\cite{Ting2026PayneZero}.

These works establish abundance-aware label-to-flux prediction as a relevant baseline. \texttt{CondGen} also implements a deterministic label-to-flux map. Its present use emphasizes forward fidelity to synthetic spectra under five elemental abundance conditions, rather than an implemented inverse-fitting pipeline. This difference in application does not make the network a stochastic sampler or imply a fundamental architectural distinction from deterministic emulators.

\subsection{Generative and physics-embedded spectral synthesis}

A smaller literature builds neural generators that synthesize spectra from labels (or latents) rather than only emulating them for fitting. Gebran~\cite{Gebran2024Astronomy} combined an autoencoder with a dense parameter-to-latent map to regenerate Gaia RVS synthetic spectra (approximately $8400$--$8800$\,\AA) from global stellar parameters including bulk metallicity $[\mathrm{M}/\mathrm{H}]$. Gebran \& Bentley~\cite{GebranBentley2025Astronomy} replaced that two-stage stack with a conditional variational autoencoder (CVAE) over an optical window (approximately $4450$--$5400$\,\AA), still conditioning on global parameters, including \Teff, $\log g$, \vsini, $[\mathrm{M}/\mathrm{H}]$, microturbulence, and resolving power~$R$, and explicitly leaving individualized elemental abundances as future work.

Physics-informed generative directions push beyond purely data-driven synthesis: PhysFormer embeds radiative-transfer / flux-conservation structure in the generative mechanism itself~\cite{Wang2026PhysFormer}, while PINN atmosphere models enforce hydrostatic (and related) constraints in a differentiable setting~\cite{Li2025PINNAtmospheres}. Score-based diffusion has begun to appear for posterior stellar spectrum sampling in exoplanet contexts through POSTELLAR~\cite{Doshi2026POSTELLAR}. Domain-translating GANs for template libraries occupy an adjacent generative niche but target observed realism rather than \texttt{SYNSPEC} fidelity~\cite{Cai2024SOST}.

We extend our earlier spectral-generation work along the \emph{abundance axis}. We summarize the differences between our previous generative \texttt{SYNSPEC} surrogates and the present \texttt{CondGen} study in Table~\ref{tab:ae-cvae-CondGen}.

\GEBRANwidefloatstart
\begin{table*}[!ht]
\caption{Contrast among our generative spectra surrogates and the present \texttt{CondGen} study. }
\label{tab:ae-cvae-CondGen}
\centering
\small
\setlength{\tabcolsep}{4pt}
\begin{tabular}{@{}>{\raggedright\arraybackslash}p{0.14\linewidth}>{\raggedright\arraybackslash}p{0.26\linewidth}>{\raggedright\arraybackslash}p{0.26\linewidth}>{\raggedright\arraybackslash}p{0.26\linewidth}@{}}
\toprule
Axis & Gebran (2024) AE \cite{Gebran2024Astronomy} & Gebran \& Bentley (2025) CVAE \cite{GebranBentley2025Astronomy} & This work (\texttt{CondGen}) \\
\midrule
Wavelength & Gaia RVS $\sim$8400--8800\,\AA & Optical $\sim$4450--5400\,\AA & Optical $\sim$4000--5400\,\AA \\
\addlinespace
Conditioning & Global $[\mathrm{M}/\mathrm{H}]$ (+ Teff, $\log g$, $v\sin i$, $\xi_t$) & Global $[\mathrm{M}/\mathrm{H}]$ (+ Teff, $\log g$, $v\sin i$, $\xi_t$); \textbf{$R$ free} & Elemental abundances Mg, Sc, Ti, Cr, Fe (+ atmospheric labels); \textbf{$R$ fixed} \\
\addlinespace
Generative form & Autoencoder + dense parameter$\to$latent map & Conditional VAE (CVAE) & Deterministic dense label-to-spectrum map; no latent sampling \\
\addlinespace
Domain / task & Synthetic \texttt{SYNSPEC} regeneration & Synthetic \texttt{SYNSPEC} regeneration & Synthetic-only forward fidelity; inverse recovery shelved \\
\bottomrule
\end{tabular}
\end{table*}
\GEBRANwidefloatend

The contribution is an extension of our existing spectral-generation program to individual abundances at fixed resolving power, used here as a research-workflow case study. No priority claim for neural spectral generation is made. The model is data-driven; it does not implement the physics-embedded mechanism of PhysFormer or the diffusion sampling of POSTELLAR.

\subsection{Foundation and transfer models}

Self-supervised and foundation-style spectral models grew rapidly in 2023--2026. Transformer foundation models for stars support multi-survey representation learning, parameter inference, and spectrum generation / inpainting~\cite{LeungBovy2024Foundation}. SpectraFM pretrains transformers on large synthetic APOGEE-like grids and transfers toward observed spectra and few-shot abundances~\cite{KoblischkeBovy2024SpectraFM}. SpecCLIP aligns and translates spectroscopic measurements across instruments (e.g.\ LAMOST LRS to and from Gaia XP) with abundance-relevant downstream hooks~\cite{Zhao2026SpecCLIP}; LoRA fine-tuning adapts such models across surveys with modest compute~\cite{Zhao2025LoRASpecCLIP}. OmniSpectra targets native-resolution multi-survey unification~\cite{Islam2026OmniSpectra}. Cross-modal galaxy image--spectrum models such as AstroCLIP illustrate related architectural ideas outside the stellar \texttt{CondGen} claim~\cite{Parker2024AstroCLIP}.

We cite this literature only as landscape. The present \texttt{CondGen} is not a foundation model, not contrastive pretraining at SpecCLIP/SpectraFM scale, and not a cross-survey transfer product. 

\subsection{Domain adaptation: deliberate out of scope}

Unsupervised and GAN-based bridges between synthetic and observed spectra are mature: Cycle-StarNet remains a canonical cycle domain-adaptation approach between theoretical and survey spectra~\cite{OBriain2021CycleStarNet}, and SOST constructs generative-adversarial template libraries that translate toward observed LAMOST-like realism~\cite{Cai2024SOST}. Those goals are scientifically important but orthogonal to a synthetic-only \texttt{CondGen} fidelity study. We therefore cite them once as a deliberate scope boundary.

\subsection{Positioning of this work}

In summary, abundance-aware \emph{deterministic} emulators (Payne lineage) and survey UQ tools such as conditional invertible networks~\cite{Candebat2024cINN,Ksoll2026cINN4MOST} define the fitting/inference frontier; foundation models define the transfer frontier; Cycle-StarNet/SOST define the domain-adaptation frontier; PhysFormer and POSTELLAR define physics-embedded and diffusion generative frontiers. Against that backdrop, this paper's contribution is tightly scoped: an abundance-conditioned \texttt{CondGen} for optical stellar spectra (Mg, Sc, Ti, Cr, Fe and related labels), trained in the synthetic domain at fixed~$R$, with forward fidelity as the primary scientific product of the case study. It extends Gebran \& Bentley~\cite{GebranBentley2025Astronomy} from global $[\mathrm{M}/\mathrm{H}]$ to elemental abundance axes, and it is architecturally and wavelength-distinct from the Gaia RVS autoencoder generator of Gebran~\cite{Gebran2024Astronomy}. Inverse abundance recovery from \texttt{CondGen} spectra is shelved for a separate study and is not centered here.

\section{Database}
\label{sec:database}

The \texttt{CondGen} case study uses a database of synthetic spectra representative of A-, F-, G-, and K-type stars. The calculations follow the atmosphere and spectral-synthesis framework used in our previous studies~\cite{Gebran2024Astronomy,GebranBentley2025Astronomy}. In this framework, \texttt{ATLAS9}~\cite{Kurucz1992} provides the stellar model atmospheres, describing how temperature, pressure, and density vary with depth for a specified effective temperature, surface gravity, and chemical composition. These one-dimensional atmospheres adopt a plane-parallel geometry, hydrostatic equilibrium, and local thermodynamic equilibrium (LTE), in which the local temperature and density determine the atomic level populations. Line blanketing accounts for the combined influence of numerous absorption lines on the atmospheric structure and the redistribution of radiation. The framework also includes convective energy transport in cooler atmospheres, as described in our earlier work.

The resulting atmospheric structures are supplied to \texttt{SYNSPEC}~\cite{2017arXiv170601859H}, which solves the radiative-transfer equation to calculate the emergent spectrum over the selected wavelength interval. Continuum and spectral-line opacities determine the wavelength-dependent flux, allowing the effects of the adopted elemental abundances to be represented in the synthetic spectra. These synthetic spectra provide the reference outputs used to train and evaluate \texttt{CondGen}; no observed spectra enter training or the primary fidelity evaluation.

Each training/evaluation spectrum is the \texttt{SYNSPEC} continuum-normalized flux vector. \texttt{CondGen} learns and is scored in that normalized flux space, not in absolute flux or magnitudes. We do not apply an additional empirical continuum re-fit at train or test time.

The database is structured in the same way we did in \cite{Gebran2024Astronomy, GebranBentley2025Astronomy} and the bot has reading access to it ($\sim$45\,GB on disk). The optical window covers $4000$--$5400$\,\AA\ at $\Delta\lambda=0.05$\,\AA\ ($N_\lambda=28\,000$ wavelength points) with a resolving power $R\simeq 50\,000$. The wavelength abscissa is constructed as
\[
\lambda_i = 4000 + i\,\Delta\lambda,\qquad i=0,\ldots,N_\lambda-1,
\]
equivalently, so the last sampled wavelength is $5399.95$\,\AA. Each continuum-normalized flux vector is paired with classical atmospheric parameters and five elemental abundances. The per-spectrum dump tuple is laid out as
\begin{itemize}
\item index~0: continuum-normalized \texttt{SYNSPEC} flux (length $N_\lambda$);
\item indices~1--5: \Teff\ (K), \logg\ (dex), \vsini\ ($\mathrm{km}\,\mathrm{s}^{-1}$), \micro\ ($\mathrm{km}\,\mathrm{s}^{-1}$), and resolving power $R$ (database metadata; fixed at $R=50\,000$);
\item index~6: $[\mathrm{Mg/H}]$;
\item indices~7--10: $[\mathrm{Sc/H}]$, $[\mathrm{Ti/H}]$, $[\mathrm{Cr/H}]$, and $[\mathrm{Fe/H}]$.
\end{itemize}
The \texttt{CondGen} conditioning vector is the standardized 9-D label
\[
\begin{aligned}
c=[&\Teff,\,\logg,\,\vsini,\,\micro,\\
&\mathrm{Mg},\,\mathrm{Sc},\,\mathrm{Ti},\,\mathrm{Cr},\,\mathrm{Fe}]
\end{aligned},
\]
i.e.\ classical atmospheric parameters together with five elemental abundances $[\mathrm{Mg/H}]$, $[\mathrm{Sc/H}]$, $[\mathrm{Ti/H}]$, $[\mathrm{Cr/H}]$, and $[\mathrm{Fe/H}]$. Resolving power $R=50\,000$ is a fixed database property and is \emph{not} an input channel of $c$. Coverage logs for the Abundance \texttt{CondGen} campaign place \Teff\ in $[4000,10000]$\,K, \logg\ in $[2.00,5.00]$ dex, $v\sin i$ in $[0,250]$\,km\,s$^{-1}$, and microturbulence $\xi_t$ in $[0.5,3]$\,km\,s$^{-1}$, with each of $[\mathrm{Mg/H}]$, $[\mathrm{Sc/H}]$, $[\mathrm{Ti/H}]$, $[\mathrm{Cr/H}]$, and $[\mathrm{Fe/H}]$ spanning $[-1.50,1.50]$ dex.

The database contains exactly $N=216\,651$ spectra.  Wavelength sampling and resolving power are fixed; the varying atmospheric parameters and abundances are described in the project as randomly sampled within the stated ranges. 

Using this database, the bots froze a stratified train/validation/test partition of the full database ($N=216\,651$) at 80\%/10\%/10\% with a constant seed for reproducibility purposes. We therefore have
\begin{align*}
n_{\mathrm{train}}&=173\,325,\\
n_{\mathrm{val}}&=21\,669,\\
n_{\mathrm{test}}&=21\,657
\end{align*}
spectra. The supplied interaction summary does not include the seed value, stratification definition, or split manifest. Those execution records are needed to reproduce the reported partition exactly.

\section{Conditional Generator}
\label{sec:model}

The bot team developed the numerical implementation within the delegated workflow described in Section~\ref{sec:ai_workflow}, drawing on our previous papers and the notebooks we provided. These materials supplied the scientific background and an implementation starting point from which the bots proposed modeling choices and planned the calculations. The Orchestrator coordinated this work, the Python/Jupyter Specialist implemented and ran the numerical tasks, and the Spectroscopy Domain Critic reviewed their outputs. We set the scientific scope, specified nine inputs from the outset, and approved the proposed training and evaluation milestones.

\subsection{Architecture}
\begin{figure*}[!ht]
\centering
\includegraphics[width=\textwidth]{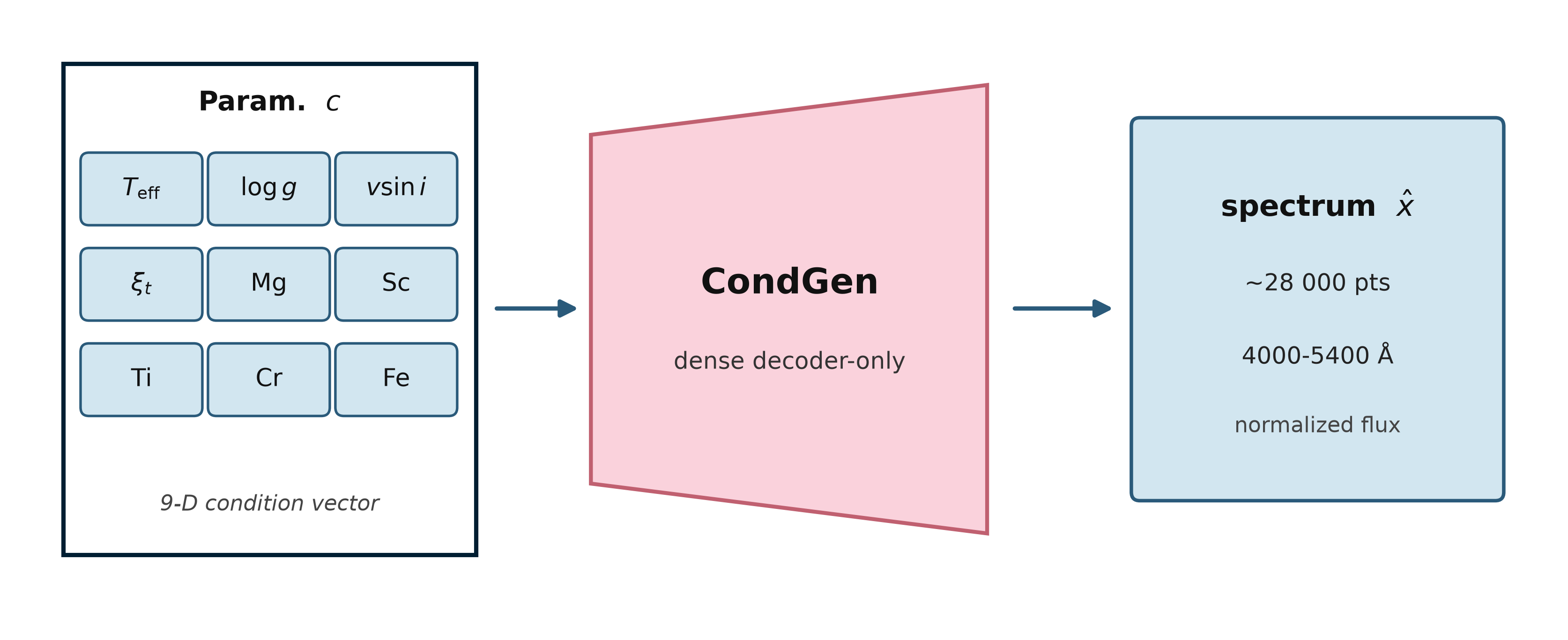}
\caption{Schematic of the 9-D \texttt{CondGen} forward model. A 9-element conditioning vector $c=[\Teff,\logg,\vsini,\micro,\mathrm{Mg},\mathrm{Sc},\mathrm{Ti},\mathrm{Cr},\mathrm{Fe}]$ is mapped by a dense decoder-only network to a normalized spectrum $\hat{x}$ over $4000$--$5400$\,\AA\ ($N_\lambda=28\,000$). Resolving power $R=50\,000$ is a fixed database property, not an input channel.}
\label{fig:arch}
\end{figure*}

\texttt{CondGen} is a dense decoder-only network that maps the standardized 9-D conditioning vector
\[
\begin{aligned}
c=[&\Teff,\,\logg,\,\vsini,\,\micro,\\
&\mathrm{Mg},\,\mathrm{Sc},\,\mathrm{Ti},\,\mathrm{Cr},\,\mathrm{Fe}]
\end{aligned}
\]

to a continuum-normalized spectrum $\hat{x}\in\mathbb{R}^{N_\lambda}$,
with $N_\lambda=28\,000$. At inference, \texttt{CondGen} produces a
single predicted spectrum for each supplied input vector, without
stochastic latent sampling, and therefore functions as a deterministic
spectral surrogate. Building on their examination of the earlier methods and notebooks, the bot team adopted the dense decoder-only architecture in place of a CVAE formulation to avoid posterior collapse. The schematic is shown in
Figure~\ref{fig:arch}.

In practice the decoder is a seven-stage fully connected stack that widens monotonically from the label vector to the spectrum. After the input layer, six hidden dense layers use widths $256$, $512$, $1024$, $2048$, $4096$, and $8192$ units, each with GELU activations and batch normalization. Dropout with rate $0.1$ is applied after the $2048$-, $4096$-, and $8192$-unit blocks. A final dense layer with $28\,000$ units and a sigmoid activation produces the continuum-normalized flux vector. There are no convolutional or recurrent layers and no stochastic latent sampling at inference: the forward map is a deterministic MLP. The reported 9-D build comprises $274\,182\,496$ parameters in total. We do not claim that this capacity is minimal. To assess alternatives, the bot team carried out the ArchSearch pilots described in Section~\ref{sec:arch_comparison}, comparing the dense baseline with a wider MLP, a FiLM residual network, a multi-scale Conv1D network, and wavelength attention under a fixed assessment protocol. The Python/Jupyter Specialist performed the comparisons and the Critic reviewed the findings. The reported pilot results supported retaining the baseline, a decision we approved; they do not establish a general ranking of these architectures.

The bots used our previous spectral-generation studies~\cite{Gebran2024Astronomy,GebranBentley2025Astronomy} as methodological references when developing the abundance-conditioned \texttt{CondGen} implementation. The 2024 model is a two-stage autoencoder plus dense parameter-to-latent map aimed at Gaia RVS wavelengths with global $[\mathrm{M}/\mathrm{H}]$. The 2025 CVAE generates optical \texttt{SYNSPEC} spectra conditioned on \Teff, $\log g$, \vsini, $[\mathrm{M}/\mathrm{H}]$, microturbulence, and resolving power~$R$. The present Methods fix~$R$ and replace bulk metallicity with elemental abundance axes (Mg, Sc, Ti, Cr, Fe), so comparisons to those priors should emphasize conditioning and wavelength/domain choices rather than recycled residual figures from 2025. Earlier AFGK and A/Am parameter-estimation work, including PCA inversion~\cite{Gebran2016AA}, regularized sliced inverse regression~\cite{Kassounian2019OpenAstronomy}, and the deep-learning label trilogy on synthetic then observed AFGK spectra~\cite{Gebran2022OpenAstronomy,Gebran2023OpenAstronomy,Gebran2025Denoising}, motivates the optical AFGK database setting and the focus on abundance-sensitive line regions, but those papers are label-inference studies rather than \texttt{CondGen} generators. When describing the forward task, Payne-style networks also implement labels to flux maps with individual abundances~\cite{Ting2019Payne,Rozanski2025TransformerPayne,Vernekar2025LRPayne}, including NLTE and data-driven survey variants~\cite{Storm2026NLTEPayne,Zhang2024DDPayne}; those systems are typically paired with $\chi^2$ or survey fitting, whereas we train a conditional generative \texttt{CondGen} and score synthetic reconstruction fidelity. Emulator scaling results~\cite{Rozanski2025ScalingLaws} inform capacity/grid-size discussion only; we do not import their published error numbers as ours. We do not embed RTE residuals or PINN atmosphere constraints in the \texttt{CondGen} loss~\cite{Wang2026PhysFormer,Li2025PINNAtmospheres}, nor do we train score-based diffusion spectrum samplers~\cite{Doshi2026POSTELLAR}; such extensions are future work.

\GEBRANwidefloatstart
\GEBRANwidefloatend

Using the supplied notebooks as a starting point, the Python/Jupyter Specialist implemented the training calculations coordinated by the Orchestrator. The documented training recipe combines mean squared error (MSE) and mean absolute error (MAE):
\[
\mathcal{L}=0.8\,\mathrm{MSE}+0.2\,\mathrm{MAE},
\]
monitored by validation MAE, with batch size $512$ and up to $150$ epochs.

\section{Scientific case-study results}
\label{sec:results}

\subsection{Forward-generation metrics and provenance}
\label{sec:forward_metrics}

The bot team used our earlier papers and the supplied notebooks to plan the forward-generation assessments reported in this section. Within the workflow described in Part~A, the Orchestrator coordinated the evaluation tasks, the Python/Jupyter Specialist performed the calculations, and the Spectroscopy Domain Critic examined the resulting diagnostics and their interpretation. The following distinctions explain how these activities relate to our published work and to the nine-input model.

\begin{itemize}
\item \textbf{Published background.} Our 2024 and 2025 studies~\cite{Gebran2024Astronomy,GebranBentley2025Astronomy} establish the scientific lineage summarized in Table~\ref{tab:ae-cvae-CondGen}. The bots drew on these studies for methodological context when developing the evaluation approach, while the supplied notebooks provided a starting point for its numerical implementation. The evaluation scores reported below come from the present project's assessments, not from those publications.
\item \textbf{Earlier project assessments.} The diagnostic files retained in this project (Figures~\ref{fig:assess_overlay_baseline}--\ref{fig:assess_abundance_sweep_Mg_baseline}) and the recorded architecture scores (Table~\ref{tab:arch_comparison}) are discussed in the Orchestrator's retrospective account as material from the baseline assessment and architecture pilot. They illustrate numerical evaluation, figure preparation, critique, and model-selection decisions within the AI-team case study. One narrative source describes these activities, but this does not establish that every output came from a single evaluation run.
\item \textbf{Nine-input model.} Figure~\ref{fig:arch} describes the design specified from the outset: four atmospheric parameters and five abundances, with resolving power fixed outside the input vector. The numerical results below are identified by their recorded assessment context, distinguishing the baseline diagnostics from the $256$-spectrum architecture pilot.
\end{itemize}

The Python/Jupyter Specialist and the Critic identified an assessment that used scalers inconsistent with training. Their diagnosis led to the use of the training-matched preprocessing protocol, and we accepted the resulting baseline MAE of $0.003681$ for reporting. This episode connects the numerical result to the bots' checking and correction process, as described in Appendix~\ref{app:workflow_scalers}.

The retrospective workflow account and the numerical summaries serve different purposes. The account describes reported assignments, checks, and decisions; the existing assessment figures and pilot table summarize numerical outputs. Their inclusion does not constitute an independent rerun or verification of the underlying calculations. The documentation accompanying these summaries does not provide a complete mapping from each output to its notebook, checkpoint, scaler files, test indices, and execution command. The results are therefore presented with their recorded sample sizes and assessment context, without treating the retrospective narrative as a substitute for execution records.

Correlation and $R^2$ values printed for individual spectra or selected subsets apply to those displayed samples and are not substituted for full-test scores. A reproducible numerical release should state each metric's aggregation and link the figures to the archived predictions, reference spectra, and sample indices.

These are synthetic forward-reconstruction diagnostics. They do not measure inverse abundance precision, agreement with observed spectra, or the efficiency of the AI-agent workflow. Numerical status and workflow outcomes are assessed separately (Table~\ref{tab:ai_outcomes}).

\subsection{Project assessment diagnostics and bot autonomy}
\label{sec:baseline_figs}

\begin{figure*}[!ht]
\centering
\captionsetup{belowskip=0pt}
\includegraphics[width=\textwidth]{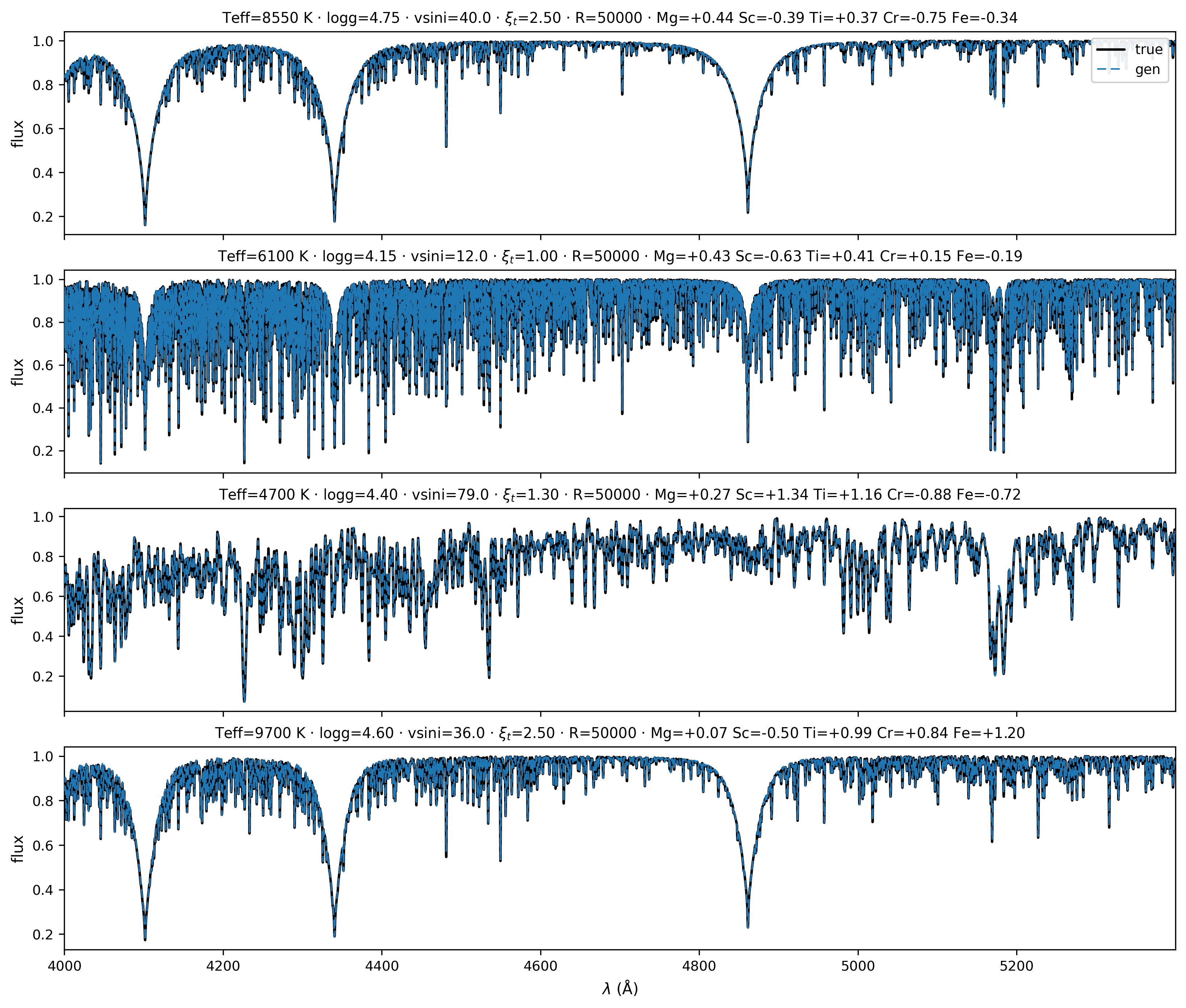}
\caption{Diagnostic from the project's baseline assessment. Representative \texttt{SYNSPEC}/true (\textbf{black solid}) versus \texttt{CondGen}-generated (\textbf{blue dashed}) spectrum overlays. Panel titles list full stellar parameters (\Teff, \logg, \vsini, \micro, $R=50\,000$, Mg, Sc, Ti, Cr, and Fe abundance).}
\label{fig:assess_overlay_baseline}
\end{figure*}

\begin{figure*}[!ht]
\centering
\captionsetup{belowskip=0pt}
\includegraphics[width=\textwidth]{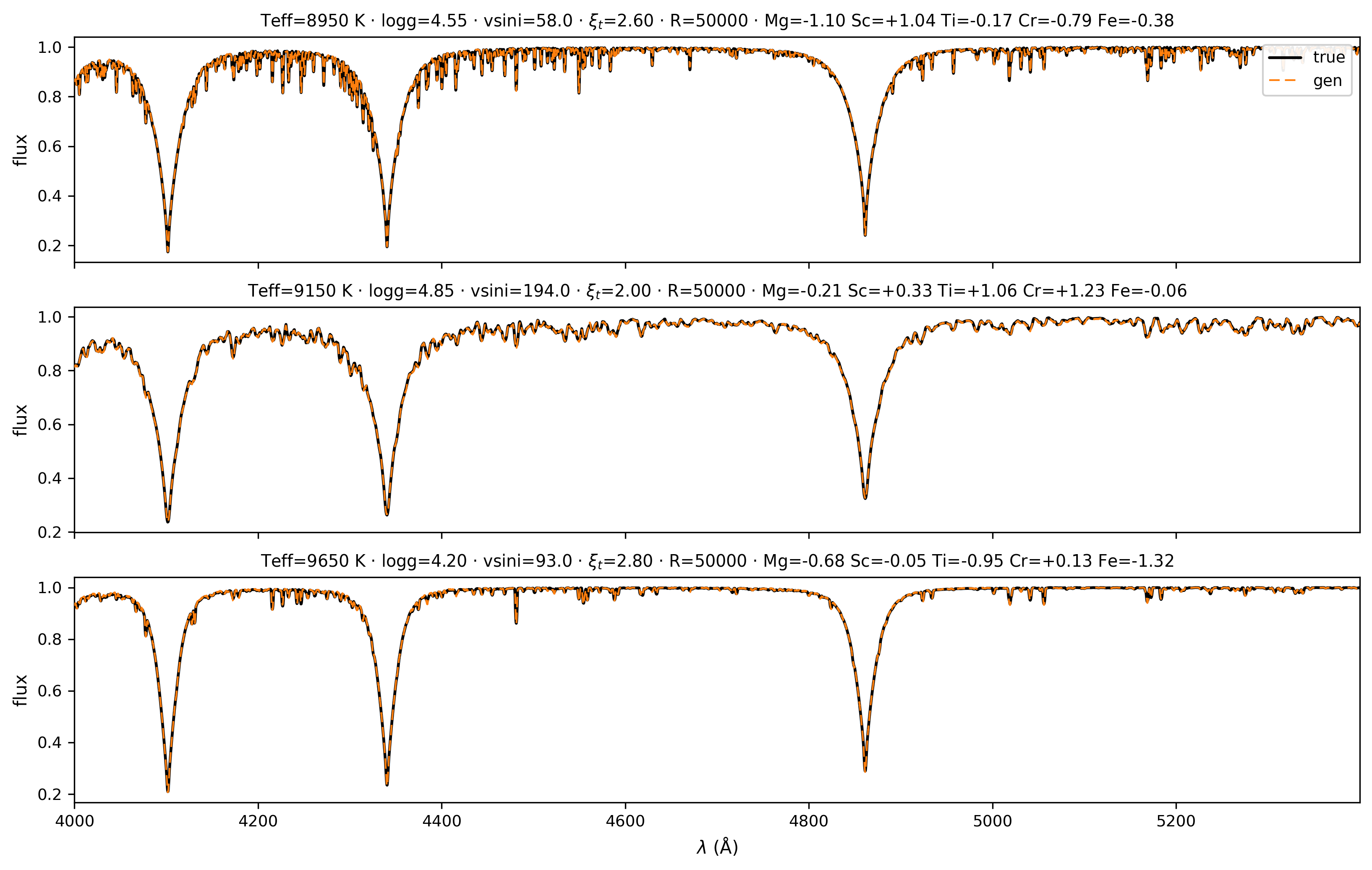}
\caption{Diagnostic from the project's baseline assessment. Best-case baseline reconstructions (qualitative panel selection by MAE). \texttt{SYNSPEC}/true (\textbf{black solid}); \texttt{CondGen} generated (\textbf{orange dashed}). Panel titles show full stellar parameters (\Teff, \logg, \vsini, \micro, $R=50\,000$, Mg, Sc, Ti, Cr, and Fe abundance).}
\label{fig:assess_best_baseline}
\end{figure*}

\begin{figure*}[!ht]
\centering
\captionsetup{belowskip=0pt}
\includegraphics[width=\textwidth]{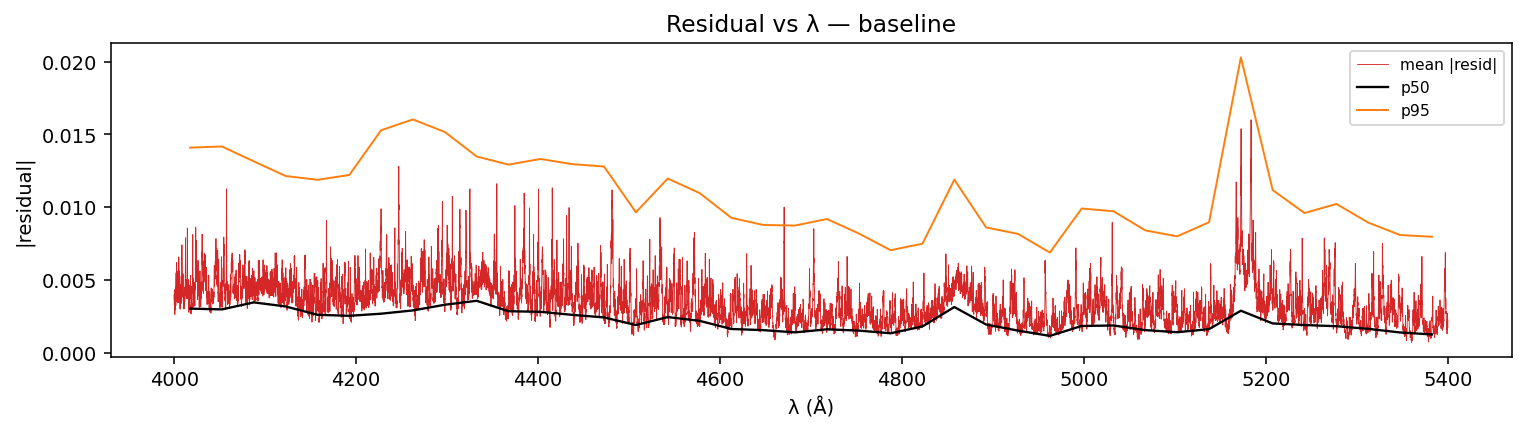}
\caption{Diagnostic from the project's baseline assessment. Absolute residuals $|\hat{x}-x|$ between \texttt{CondGen}-generated and \texttt{SYNSPEC} spectra as a function of wavelength. The curves show the mean absolute residual (red), median (p50; black), and 95th percentile (p95; orange).}
\label{fig:assess_residual_vs_lambda_baseline}
\end{figure*}

Figures~\ref{fig:assess_overlay_baseline}--\ref{fig:assess_residual_vs_lambda_baseline} illustrate how specialist bots developed an assigned evaluation task into numerical results, visual diagnostics, and manuscript material. Drawing on the supplied papers and notebooks, the Python/Jupyter Specialist carried out the reconstruction calculations and passed numerical outputs and stellar-parameter information to the Figure Artist. The Artist prepared the diagnostic panels, and the \LaTeX\ Editor incorporated them into the manuscript. The Orchestrator coordinated these handoffs, while the Spectroscopy Domain Critic challenged scientific interpretations and presentation choices, prompting revisions through exchanges among the specialists (Appendix~\ref{app:workflow_manuscript}).

The bots exercised autonomy within their assigned responsibilities: numerical execution, figure preparation, and technical revisions were handled through the coordinated specialist workflow. Outputs from one bot became inputs for another, allowing the team to advance connected research tasks and respond to critique. We defined the scientific scope, approved milestones, and reviewed the interpretation and reporting. This division of responsibilities illustrates how bot initiative and collaboration operated within human supervision.

The Orchestrator's account associates these figures with the baseline assessment. It describes the collaboration but does not provide a separate execution and review record for each panel. These figures illustrate the reported baseline diagnostics and their role in the bot-led assessment workflow. 
Figure~\ref{fig:assess_overlay_baseline} shows representative \texttt{SYNSPEC}/true (\textbf{black solid}) and \texttt{CondGen}-generated (\textbf{blue dashed}) overlays. The Python-Artist handoff connected calculated spectra to labeled visual comparisons, making agreement and local mismatches available for scientific inspection. This illustrates how one bot's numerical work supported another bot's preparation of a research output.

\GEBRANwidefloatstart
\clearpage
\GEBRANwidefloatend

Figure~\ref{fig:assess_best_baseline} shows selected best-case examples (\texttt{SYNSPEC}/true \textbf{black solid}; \texttt{CondGen} \textbf{orange dashed}). Here, numerical assessment and figure preparation are connected through selection by MAE, illustrating how the bots translated computed results into examples for the manuscript. These selected reconstructions demonstrate close agreement in particular cases; they do not estimate average performance across the test set.

\GEBRANwidefloatstart
\GEBRANwidefloatend

The Python/Jupyter Specialist computed the differences between generated and reference spectra, and the Figure Artist plotted their absolute values against wavelength in Figure~\ref{fig:assess_residual_vs_lambda_baseline}. The purpose of this bot-produced diagnostic was to examine where reconstruction errors are concentrated across the spectrum. The plot shows the mean absolute residual, the median (p50), and the 95th percentile (p95). Errors are generally larger in the $4000$-$4500$\,\AA\ region and lower across much of $4600$-$4800$\,\AA, with a pronounced local enhancement near $5170$\,\AA. The contrast between larger deviations in regions containing strong or crowded absorption features and smaller deviations in quieter intervals shows that reconstruction quality varies with spectral structure. The separation between the median and the 95th percentile also highlights the spread in residual magnitudes, with larger errors extending beyond the typical level. This diagnostic therefore gave the bot team a way to locate regions requiring closer inspection and supplied the Critic with material for assessing the model's limitations, complementing the selected examples of close agreement.

\GEBRANwidefloatstart
\GEBRANwidefloatend

\begin{figure*}[!ht]
\centering
\captionsetup{belowskip=0pt}
\includegraphics[width=\textwidth]{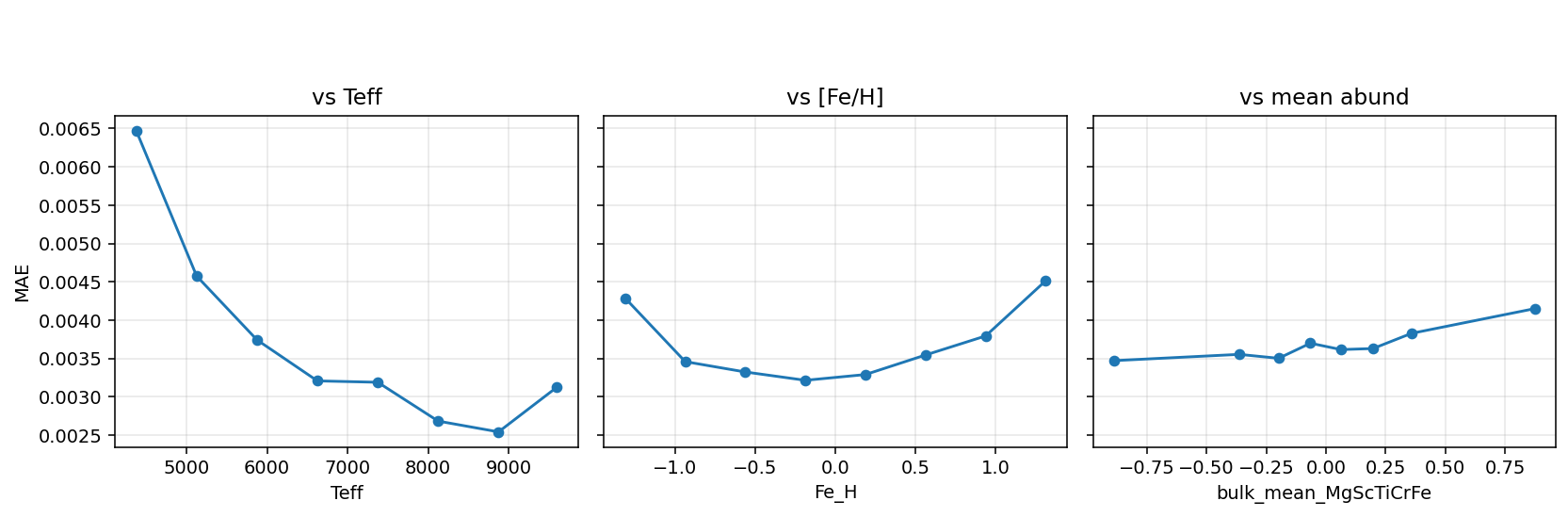}
\caption{Existing full-test ($n=21\,657$) binned MAE summaries for the \texttt{CondGen} baseline versus \Teff\ (left), $[\mathrm{Fe/H}]$ (middle), and the mean of the five elemental abundance labels in dex (right). Each panel summarizes variation across the remaining labels. Only MAE is plotted; bin counts, uncertainty intervals, and binned RMSE are not shown.}
\label{fig:binned_mae_teff_fe_bulk}
\end{figure*}

The reported global test-set reconstruction MAE is $0.003681$ for $n=21\,657$ spectra, with reported RMSE $0.009794$, correlation $0.9979$, and $R^2=0.9958$. This average across continuum-normalized spectra does not describe every region of the label space equally. Figure~\ref{fig:binned_mae_teff_fe_bulk} presents the existing full-test binned MAE summaries versus \Teff, $[\mathrm{Fe/H}]$, and the mean of the five supplied elemental abundance labels. The latter quantity is an arithmetic average in dex, $\overline{A}=\frac{1}{5}\sum_{X\in\{\mathrm{Mg,Sc,Ti,Cr,Fe}\}}[\mathrm{X/H}]$, used as a descriptive abundance index rather than an independently specified atmospheric bulk metallicity. The reported MAE is approximately $0.00647$ in the $4000$--$4750$\,K bin and $0.00254$ in the $8500$--$9250$\,K bin. The abundance panels also show non-uniform errors. These are marginal summaries over the other labels, rather than controlled estimates of the effect of changing one physical parameter.

The displayed curves contain MAE only; no binned RMSE curves or uncertainty intervals are presented. The plot does not report per-bin sample counts or a complete numerical bin table, so differences between sparsely and densely populated bins cannot be assessed from it alone. Moreover, each full-spectrum MAE includes continuum pixels as well as absorption features. Binning by stellar parameters therefore addresses variation across the test population but does not establish line-core accuracy or remove continuum weighting. Figure~\ref{fig:assess_residual_vs_lambda_baseline} supplies complementary wavelength-dependent diagnostics. Together, these results support a descriptive assessment of reconstruction behavior, without establishing uniform accuracy across all spectral features or atmospheric conditions.

\GEBRANwidefloatstart
\clearpage
\GEBRANwidefloatend

\subsection{Abundance sweeps (Sc, Ti, Cr, Fe) and cool-star Mg-line response}
\label{sec:abundance_sweeps}

The bot team used one-element abundance sweeps to examine how \texttt{CondGen} responds when the chemical composition changes. The Python/Jupyter Specialist varied one abundance while holding the atmospheric parameters and other abundances fixed, and the Figure Artist presented the resulting spectra and line diagnostics. This controlled calculation allowed the Critic to examine whether increasing an abundance produced a consistent change in the corresponding absorption features, and to identify departures requiring further investigation.

The Orchestrator's account describes favorable qualitative assessments of the selected Sc, Ti, and Cr responses and a caveat for Fe. The bots also identified a reversal in the cool-star Mg response and brought it to the Critic's attention. We retained that diagnostic as a potential problem requiring a matched reference comparison (Appendix~\ref{app:workflow_abundances}). The plotted sweeps show \texttt{CondGen} outputs alone. Their qualitative trends can identify questions for further investigation, but do not by themselves establish agreement or disagreement with \texttt{SYNSPEC} across the swept abundances.

All five sweeps use the atmospheric anchor $\Teff=5150$\,K, $\log g=4.45$, $\vsini=78.0$\,km\,s$^{-1}$, $\micro=0.90$\,km\,s$^{-1}$, and $R=50\,000$. In each sweep, the four elements that are not varied retain their reference abundances: Mg~$=+0.67$, Sc~$=-1.38$, Ti~$=+0.49$, Cr~$=-1.28$, and Fe~$=+0.12$ in dex $[\mathrm{X/H}]$. Colors run from dark purple at the lowest swept abundance to yellow or yellow-green at the highest.

Each figure presents the diagnostic in three steps: the full generated spectrum, a zoom around a selected absorption feature, and the mean flux in the selected wavelength window as a function of abundance. A decrease in mean normalized flux indicates stronger absorption within that window. The bots could therefore inspect both the changing line profiles and a compact measure of their response. These checks assess qualitative behavior at one atmospheric anchor; they do not establish a numerical accuracy threshold across the full label space.

\begin{figure*}[!ht]
\centering
\captionsetup{belowskip=0pt}
\includegraphics[width=\textwidth]{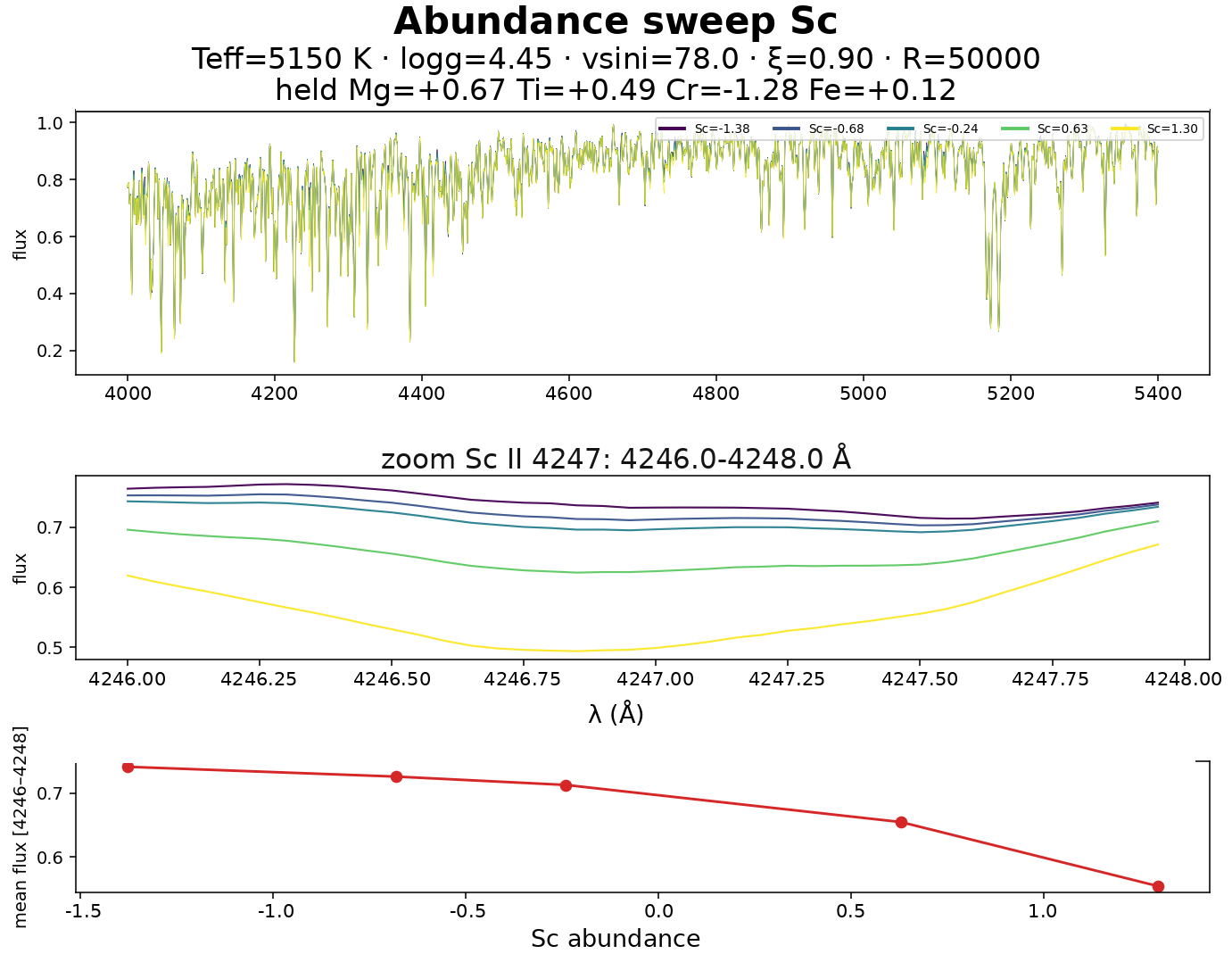}
\caption{Baseline Sc abundance sweep at the common anchor, showing the Sc\,II feature around $4247$\,\AA\ and the decrease in mean flux as abundance rises.}
\label{fig:assess_abundance_sweep_Sc_baseline}
\end{figure*}

For Sc (Figure~\ref{fig:assess_abundance_sweep_Sc_baseline}), the bots examined the feature around $4247$\,\AA. Increasing $[\mathrm{Sc/H}]$ lowers the flux in the selected window and deepens the absorption profile, while the mean-flux curve decreases steadily across the sampled abundances. The diagnostic therefore shows a consistent strengthening of the selected feature as Sc abundance increases, supporting the Critic's favorable qualitative assessment at this atmospheric anchor.

\GEBRANwidefloatstart
\GEBRANwidefloatend

\begin{figure*}[!ht]
\centering
\captionsetup{belowskip=0pt}
\includegraphics[width=\textwidth]{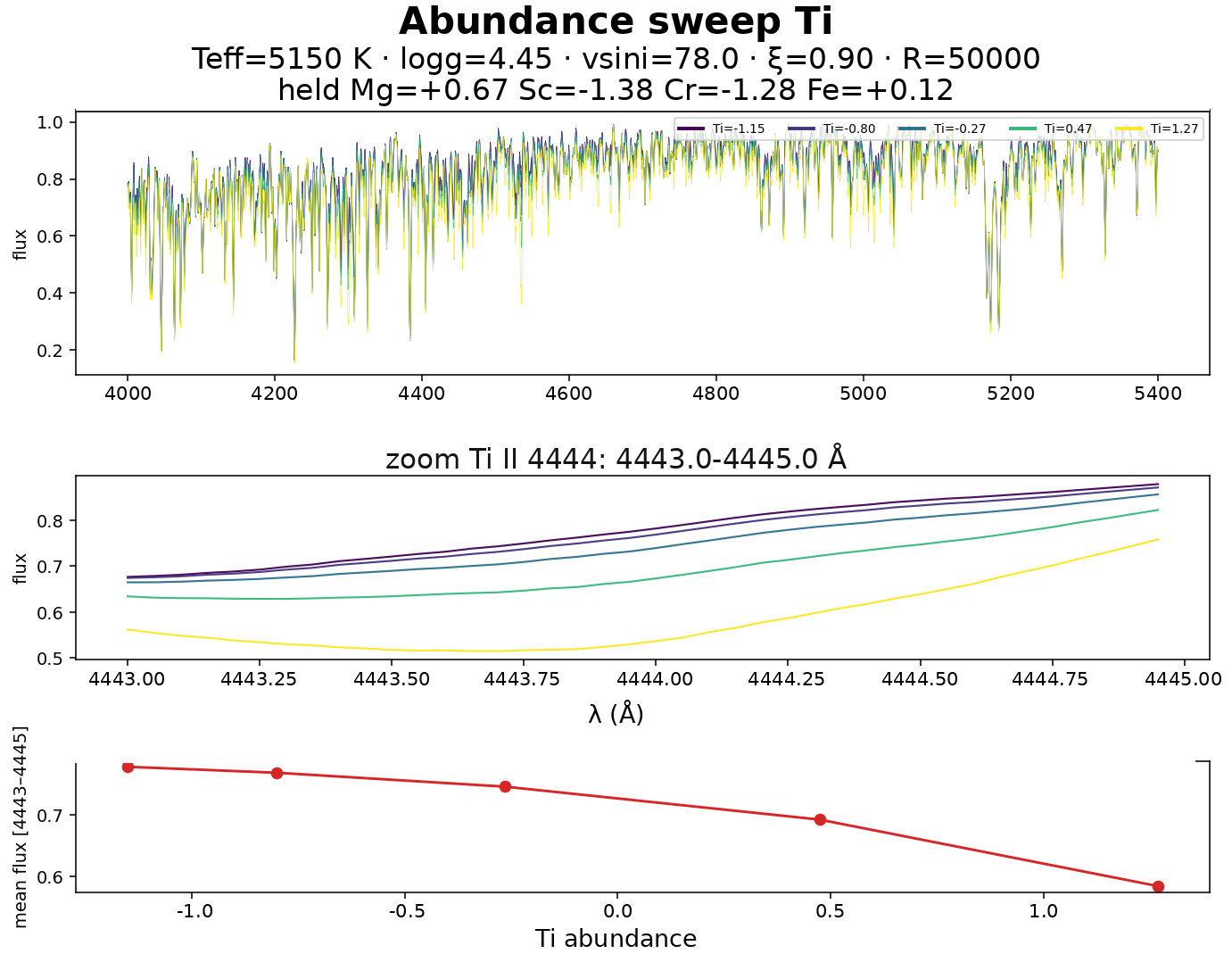}
\caption{Baseline Ti abundance sweep at the common anchor, showing the Ti\,II region around $4444$\,\AA\ and its decreasing mean-flux response.}
\label{fig:assess_abundance_sweep_Ti_baseline}
\end{figure*}

For Ti (Figure~\ref{fig:assess_abundance_sweep_Ti_baseline}), the bots plotted the response around $4444$\,\AA\ to test the effect of changing $[\mathrm{Ti/H}]$. The higher-abundance curves show deeper absorption, and the mean flux decreases consistently as Ti abundance rises. This output gave the Critic a second example of an orderly abundance response, supporting the reported qualitative acceptance of the Ti diagnostic.

\GEBRANwidefloatstart
\clearpage
\GEBRANwidefloatend

\begin{figure*}[!ht]
\centering
\captionsetup{belowskip=0pt}
\includegraphics[width=\textwidth]{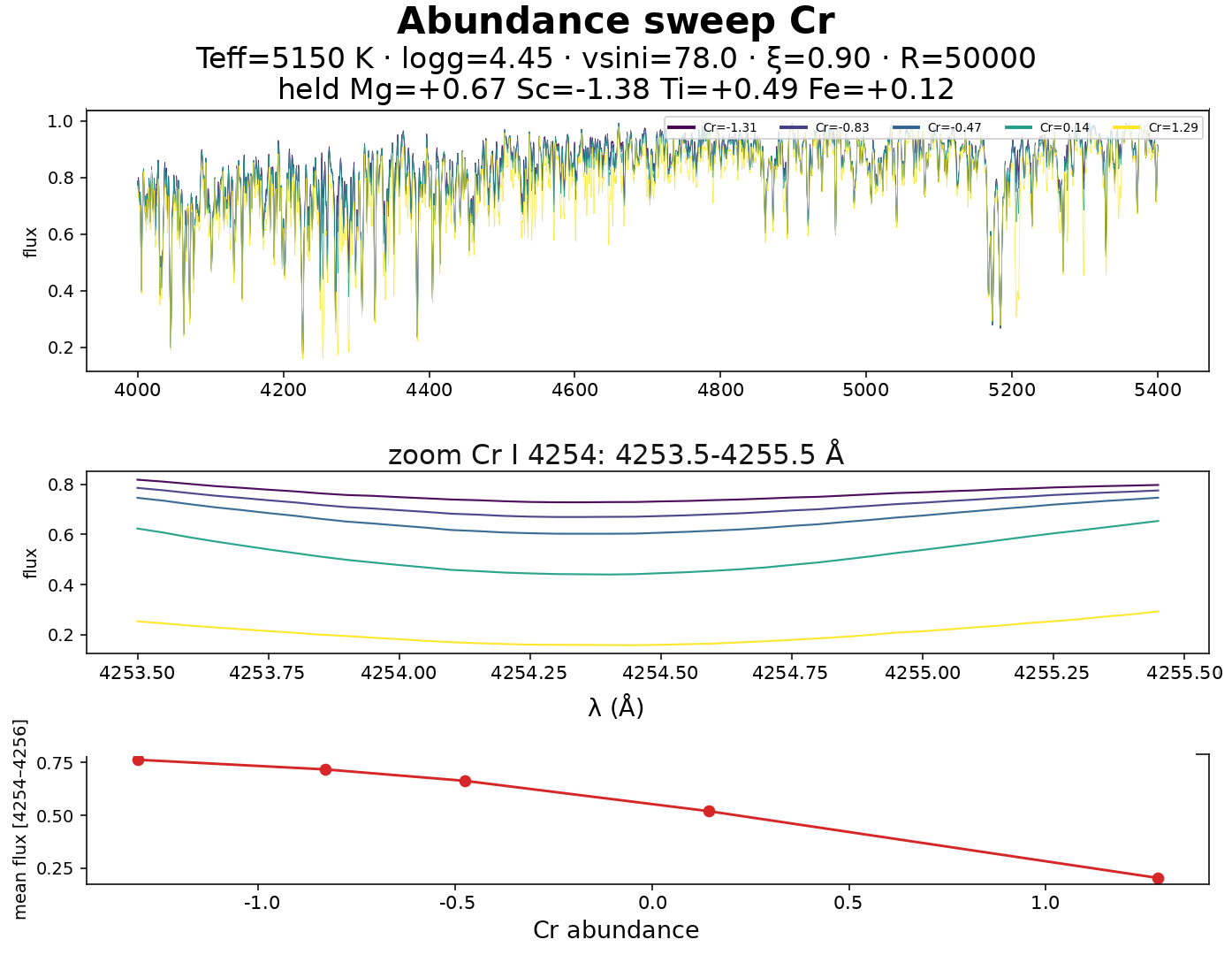}
\caption{Baseline Cr abundance sweep at the common anchor, showing the Cr\,I feature around $4254$\,\AA\ and its decreasing mean-flux response.}
\label{fig:assess_abundance_sweep_Cr_baseline}
\end{figure*}

For Cr (Figure~\ref{fig:assess_abundance_sweep_Cr_baseline}), the bots examined the feature around $4254$\,\AA. The profiles become progressively deeper as $[\mathrm{Cr/H}]$ increases, with a corresponding decline in the mean flux. The highest-abundance curve shows the strongest absorption in the displayed sequence. This diagnostic completes the three elemental responses for which the Critic reported favorable qualitative behavior, linking the numerical sweep to an explicit scientific assessment.

\GEBRANwidefloatstart
\GEBRANwidefloatend

\begin{figure*}[!ht]
\centering
\captionsetup{belowskip=0pt}
\includegraphics[width=\textwidth]{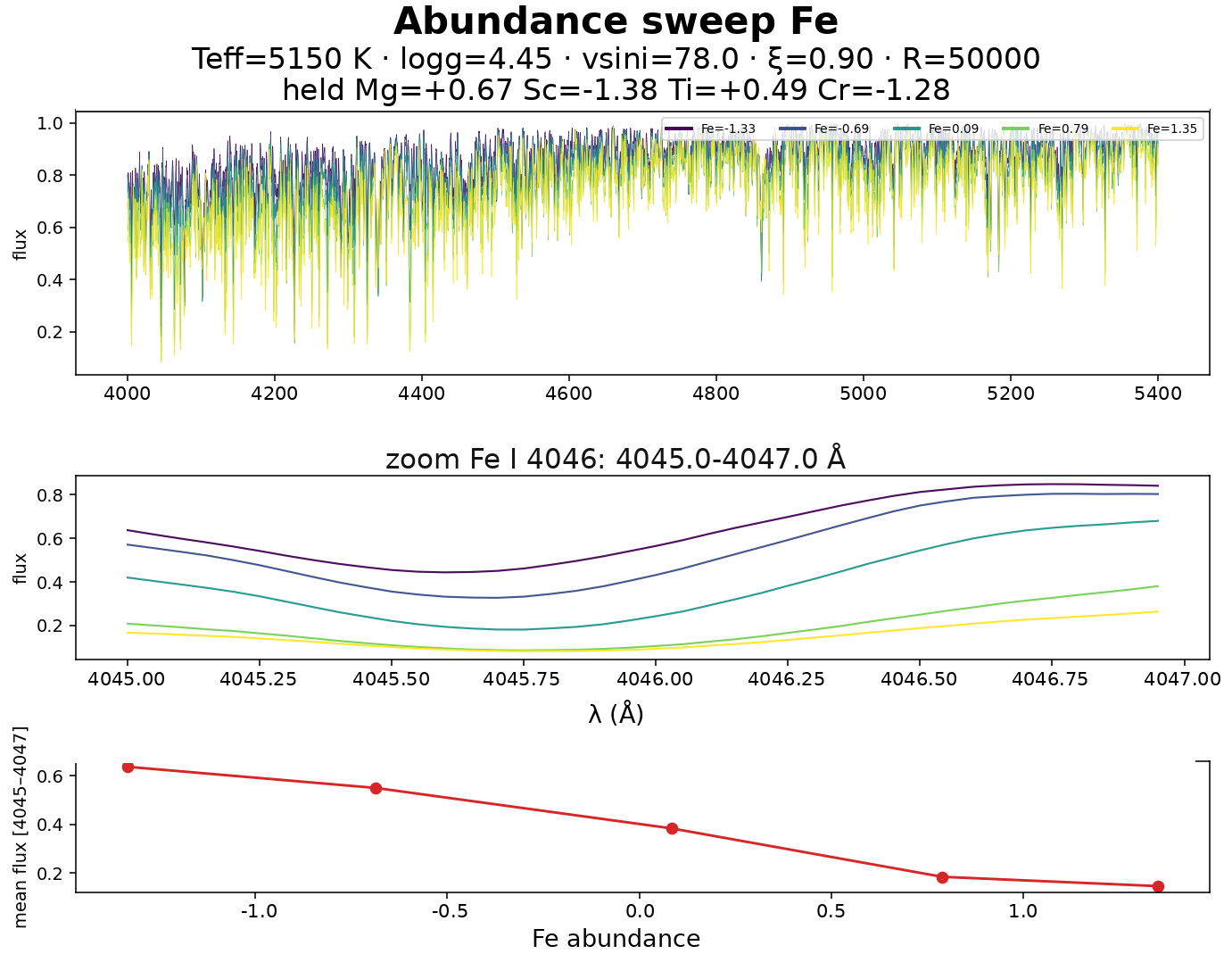}
\caption{Baseline Fe abundance sweep at the common anchor. The mean-flux response around the Fe\,I $4046$\,\AA\ feature flattens toward the highest abundances.}
\label{fig:assess_abundance_sweep_Fe_baseline}
\end{figure*}

For Fe (Figure~\ref{fig:assess_abundance_sweep_Fe_baseline}), the bots inspected the feature around $4046$\,\AA. Absorption strengthens as $[\mathrm{Fe/H}]$ rises, while the mean-flux decrease becomes smaller between the two highest abundances. The Critic identified this flattening as a feature to investigate. The generated curves alone do not establish whether the response differs from a matched \texttt{SYNSPEC} sweep.

\GEBRANwidefloatstart
\GEBRANwidefloatend

\begin{figure*}[!ht]
\centering
\captionsetup{belowskip=0pt}
\includegraphics[width=\textwidth]{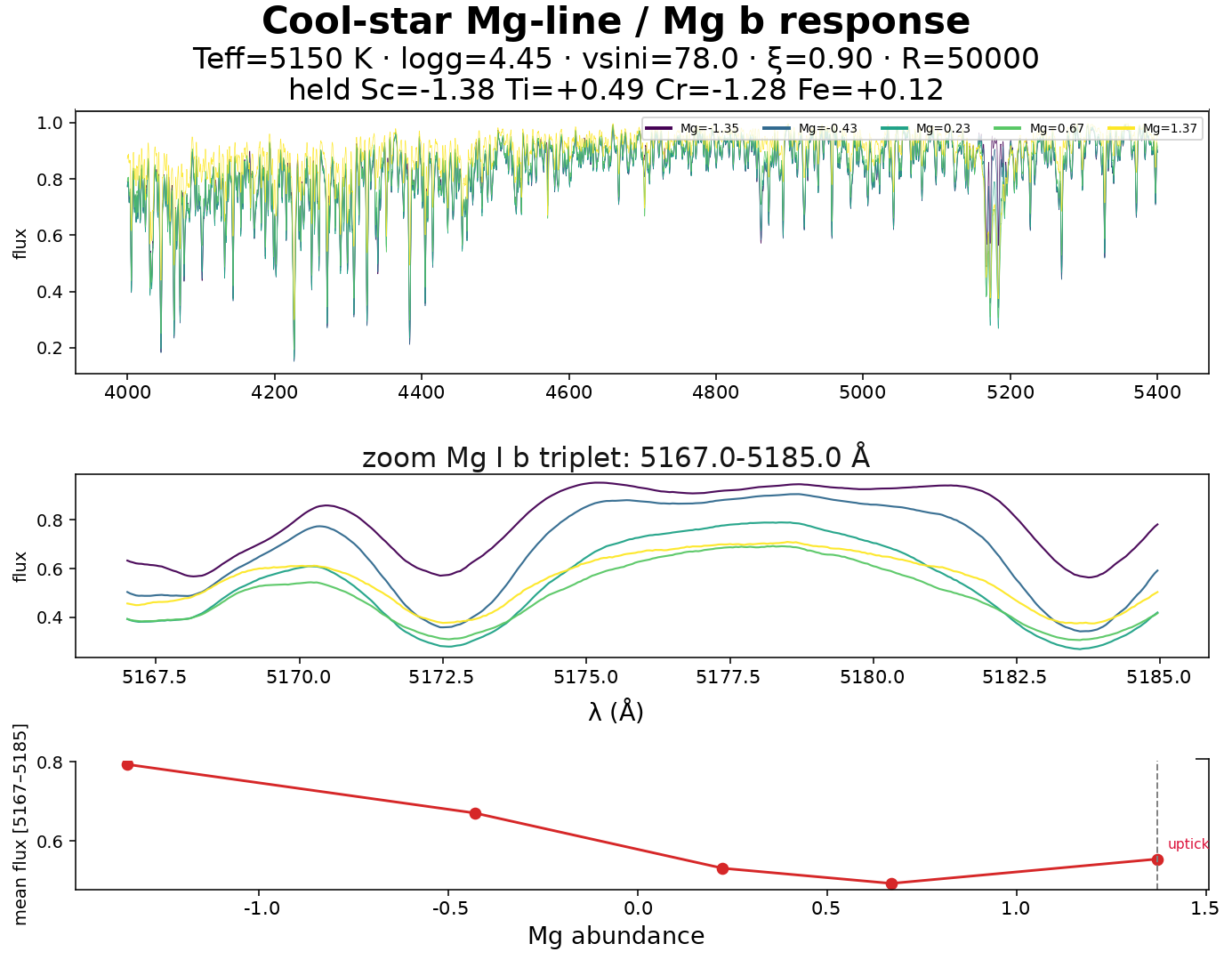}
\caption{Cool-star Mg abundance sweep generated by \texttt{CondGen}. The rise in mean flux between approximately $+0.67$ and $+1.37$\,dex is the non-monotonic response flagged by the bots. A matched \texttt{SYNSPEC} sweep is not available, so the plot identifies a potential problem for investigation rather than demonstrating a physical-model discrepancy.}
\label{fig:assess_abundance_sweep_Mg_baseline}
\end{figure*}

For Mg (Figure~\ref{fig:assess_abundance_sweep_Mg_baseline}), the bots examined the Mg\,b region over approximately $5167$--$5185$\,\AA. The generated absorption initially strengthens as $[\mathrm{Mg/H}]$ rises. At approximately $+1.37$\,dex, the mean flux rises relative to the $+0.67$\,dex case and the profiles become shallower. The bots identified this reversal and the Critic flagged it for further investigation. It is an observed property of the network output, not a demonstrated discrepancy with physical synthesis. At this fixed atmosphere, the database is reported to contain an exact \texttt{SYNSPEC} spectrum only for Mg~$=+0.67$\,dex (id~113144). The additional synthesis calculations needed for a matched sweep were not performed; no reference curves are therefore shown. Establishing the physical fidelity of the response requires that comparison with the other labels, broadening, resolution, and normalization held consistent.

\GEBRANwidefloatstart
\GEBRANwidefloatend

The Mg episode illustrates a specific contribution of the bot team: detecting an unexpected response, questioning its interpretation, and identifying the reference calculation needed to evaluate it. The workflow outcome is the identification and reporting of an open scientific question; the present figure does not establish its physical cause or demonstrate a discrepancy with a matched \texttt{SYNSPEC} sweep.

\subsection{Architecture comparison (short pilot)}
\label{sec:arch_comparison}

The bot team carried out a short architecture comparison to test whether a different network design could improve spectral reconstruction. The Python/Jupyter Specialist performed the numerical comparisons, the Critic assessed the reported outcomes, and the Orchestrator coordinated the work and communicated the findings. This gave the team a basis for model selection within the project, with us reviewing and approving the resulting choice (Appendix~\ref{app:workflow_architectures}). The bots compared the dense baseline with four alternatives, each testing a different modeling idea:

\begin{itemize}
\item \textbf{Wider MLP:} a larger fully connected decoder, used to test whether additional network capacity improved the mapping from stellar labels to spectra.
\item \textbf{Wavelength attention:} a decoder using attention across wavelengths, used to test whether this mechanism improved the representation of spectral features.
\item \textbf{Multi-scale Conv1D:} a decoder using one-dimensional convolutional filters of different sizes to capture local line structure and broader spectral patterns.
\item \textbf{FiLM residual:} a residual network in which the conditioning labels modify intermediate features through feature-wise linear modulation (FiLM), testing another way of incorporating stellar parameters and abundances.
\end{itemize}

The project record reports that all five architectures were assessed on a pilot subset of $256$ spectra from the test partition ($\mathrm{MAX\_TEST}=256$), using a common database and fixed data splits. Table~\ref{tab:arch_comparison} retains the existing MAE estimates and intervals reported as bootstrap $95\%$ confidence intervals from $2000$ resamples. The accompanying record does not specify the resampling unit, interval construction, random seed, or exact subset-selection procedure. The intervals are reproduced as reported summaries, rather than used here to claim a confirmed paired significance test. No paired difference intervals or full-test comparison of all five architectures are presented.

The comparison describes these particular pilot runs. It does not quantify variability across repeated training runs or establish equivalence of training budgets and stopping rules. Because the pilot drew from the test partition and informed model selection, that partition was not wholly untouched by the selection process; a future controlled comparison should use validation data for selection and reserve an independent test set for final assessment. Within the present workflow case study, the scores document the information available to the bot team and the authors when retaining the baseline.

In this pilot subset, the dense baseline recorded the lowest MAE ($0.003655$; CI$_{95}$~$[0.003353,\,0.003971]$) among the configurations compared (Table~\ref{tab:arch_comparison}). Wavelength attention was the closest alternative (MAE~$0.009588$; CI$_{95}$~$[0.008962,\,0.010233]$), while the FiLM residual configuration produced the largest errors (MAE~$0.018974$; CI$_{95}$~$[0.016314,\,0.022330]$). The wider MLP and multi-scale Conv1D also did not improve on the baseline in these runs. These reported point estimates informed the decision to retain the baseline for this project, which we approved; this is a description of the pilot decision rather than a statistically established ranking of model families. The unsuccessful alternatives remain in the table so that the basis for that decision is visible.

\GEBRANwidefloatstart
\begin{table}[H]
\caption{Reported MAE and bootstrap $95\%$ confidence intervals for the $256$-spectrum architecture pilot ($2000$ resamples). Values are retained from the existing assessment summaries. The resampling procedure and subset-selection method are not fully documented in the accompanying record, and the intervals are not used to assert a paired significance test or a general architecture ranking.}
\label{tab:arch_comparison}
\centering
\begin{tabular}{lc}
\toprule
Architecture & MAE [CI$_{95}$] \\
\midrule
Baseline in the pilot & $0.003655$ [$0.003353$, $0.003971$] \\
Wavelength attention & $0.009588$ [$0.008962$, $0.010233$] \\
Wider MLP & $0.011624$ [$0.010877$, $0.012434$] \\
Multi-scale Conv1D & $0.012988$ [$0.011955$, $0.014072$] \\
FiLM residual & $0.018974$ [$0.016314$, $0.022330$] \\
\bottomrule
\end{tabular}
\end{table}
\GEBRANwidefloatend

\section{Discussion}
\label{sec:discussion}

The principal result of this case study is a documented pattern of delegated work, critique, and human decision-making. The four episodes connect the role descriptions to consequential actions: rejecting an incompatible evaluation recipe, identifying a non-monotonic Mg response through bot analysis and approving its disclosure through human review, limiting the interpretation of pilot results, and coordinating manuscript artifacts. These outcomes make the AI-team contribution concrete while exposing its dependence on existing scientific infrastructure and human supervision.

\subsection{Discussion of the AI role}

\subsubsection{What collaboration contributed}

The reported benefit of specialization was a division of responsibilities and interfaces. Python supplied numerical and parameter information; the Artist and Editor turned those outputs into figures and text; the Critic challenged interpretation; and the Orchestrator routed unresolved decisions to us. The Mg episode is especially informative: the bots identified the reversal, the Critic questioned its interpretation, and we approved reporting it as a potential problem requiring a matched reference comparison. Collaboration therefore changed what was claimed while preserving the diagnostic that exposed the problem. The scaler episode similarly shows why evaluation artifacts must be checked against training provenance before numbers enter the paper.

Human input was substantive. We set the problem, supplied the data and notebooks, specified the nine-input design from the outset, approved training, and set the manuscript focus on the AI-team contribution. Describing that involvement only as light final approval would understate the decisions summarized in Appendix~\ref{app:workflow_evidence}. The case demonstrates how researchers working together can coordinate specialized agents across connected tasks.

The authors estimate that the reported analysis, diagnostics, figure preparation, and initial manuscript drafting were assembled in less than a week through iterative human prompting, interaction with the Orchestrator, and exchanges among specialist bots. This interval began with an existing \texttt{SYNSPEC} database, notebooks, and research background; it excludes their original development and does not describe the subsequent peer-review revision cycle. We did not measure the time that a human team or a single agent would require when starting from the same resources, and elapsed time was not separated into computation, waiting, and active human effort. The experience illustrates the feasibility of rapid iteration within this project, while comparative efficiency and labor savings require a study with matched starting conditions.

\subsubsection{What the workflow did not establish}

The retrospective source is a participant account rather than a complete execution trace. It cannot establish the frequency of missed errors, the proportion of tasks completed without intervention, or the counterfactual performance of a single agent. Role separation also does not guarantee independent verification.

These limitations suggest concrete improvements to research practice: preserve task identifiers and timestamps; link every numerical claim and figure to code, data, checkpoint, scaler, and evaluation output; retain valid negative results; and record human decisions alongside the agent proposals that prompted them. A prospective study could then compare configurations using task completion, independently checked error rates, human effort, runtime, and cost. Such comparisons were not performed here.

\subsection{Discussion of the science case}

The \texttt{CondGen} case study represents an important step in our research on stellar spectral generation, extending the conditioning from global metallicity toward individual elemental abundances at fixed resolving power~\cite{Gebran2024Astronomy,GebranBentley2025Astronomy}. Its nine-input design combines four atmospheric parameters with the abundances of Mg, Sc, Ti, Cr, and Fe, providing a framework for investigating how changes in chemical composition affect generated spectra. As a deterministic forward surrogate, it shares the label-to-flux formulation of established spectral emulators~\cite{Ting2019Payne,Vernekar2025LRPayne,Rozanski2025TransformerPayne}. Its value here lies in developing this abundance-conditioned case study through a coordinated AI research workflow.

The reported baseline assessments provide concrete scientific guidance for further development. The bots produced spectral reconstructions, examined the wavelength dependence of their errors, varied individual abundances, and compared alternative architectures. The Sc, Ti, and Cr sweeps showed consistent strengthening of the selected absorption features, while the Fe and Mg diagnostics identified responses that deserve closer investigation. In particular, the non-monotonic cool-star Mg response discovered by the bots defines a specific research question: does the reversal persist across atmospheric conditions, does it differ from matched synthetic references, and which changes improve agreement? These findings turn the case study into a focused program of testable questions, with the scope of the reported assessments specified in Section~\ref{sec:forward_metrics}.

A more detailed study could build directly on these results by extending the diagnostic sweeps across temperature, surface gravity, and abundance, and by testing how training coverage and model design affect the regions with the largest residuals. Targeted Mg experiments could examine the contributions of strong-line behavior, blending, and the learned spectral mapping. Architecture comparisons with matched training budgets and independent evaluation data would help identify which changes improve reconstruction accuracy. The current work supplies the scientific motivation and initial diagnostics for these investigations, supporting a practical route toward more reliable abundance-conditioned spectral generation.

The project also provides a feasible foundation for sustained research development. The existing spectral database and notebooks can support successive experiments, while the division of work among computation, scientific critique, visualization, and writing provides an organized way to carry those experiments through to interpretation and communication. Under our direction, the Orchestrator could coordinate these follow-up tasks and the specialist bots could implement, assess, and document the resulting improvements. When it identifies a need for additional expertise or capacity, the Orchestrator could create new specialist bots, define their responsibilities, and integrate their contributions into the existing team. The team could thus evolve in response to new research questions and issues raised during analysis or review. This combination of reusable resources, concrete scientific questions, and coordinated AI contributions makes the study a valuable step toward a continuing research program in abundance-conditioned spectral generation.

\section{Conclusion}
\label{sec:conclusion}

This work demonstrates the practical value of organizing specialized AI bots as a research team under human scientific direction. In the \texttt{CondGen} case study, the bots contributed numerical implementation, spectral diagnostics, architecture comparisons, scientific critique, visualization, and manuscript preparation. Their interactions helped identify an inconsistent evaluation procedure and a non-monotonic cool-star Mg response requiring reference validation, while we guided the research questions, assessed the interpretations, and approved the reporting. The main contribution is a concrete example of how human judgment and coordinated bot activity can support connected stages of a scientific investigation.

This experience motivates investigation of whether coordinated AI assistance can shorten the cycle between a question, its computational investigation, critical assessment, and the next experiment. Starting with an existing \texttt{SYNSPEC} database and notebooks, the authors estimate that the reported analysis, diagnostics, figures, and initial manuscript draft were assembled in less than a week. This account demonstrates a practical sequence of work, while a comparison using identical starting resources and recorded human effort would be needed to establish any efficiency gain. Future studies could examine whether such workflows give researchers greater capacity to explore alternatives and follow up unexpected findings.

The case presented here is a starting point for more complex forms of human-bot collaboration. Future projects could involve larger numbers of specialist bots and several human researchers, each supported by a team of agents. Bots could work independently within defined assignments, while their Orchestrators coordinate the exchange of data, code, and findings across teams. One group could develop models, another examine their statistical robustness, and another attempt to reproduce the results. The human researchers would bring these contributions together, compare interpretations, and set the shared scientific direction. When new questions or gaps in expertise emerge, the Orchestrators could create additional specialist bots and integrate them into the collaboration. This would allow the organization of the research team to evolve with the investigation.

Although this case study was carried out using Grok Bot, the approach is not inherently tied to that platform or to a single AI provider. The principles of task delegation, specialist collaboration, critique, and human supervision can be adapted to other suitable AI systems. Future improvements in AI algorithms, agent-coordination tools, and integration with scientific software could make these teams easier to establish and manage. Such developments could reduce the practical effort required to organize human--bot collaboration and make increasingly complex research projects accessible to a wider range of research groups.

For our case study application, this approach offers a practical route toward broader abundance-response studies, a matched reference investigation of the Mg response, and extensions from synthetic to observed stellar spectra. Coordinated bot teams could prepare observational data, investigate differences between simulated and real spectra, implement candidate methods, and evaluate their performance through shared diagnostics and scientific review. These tasks would extend the present case into a more demanding research program while reusing its data, numerical tools, and division of responsibilities.

Realizing this broader vision requires clear task ownership, accessible experiment records, and reproducible checks of the results. Prospective studies should examine how different team sizes and coordination strategies affect reliability, human effort, elapsed time, and cost. The wider opportunity is a research environment in which human insight and bot autonomy reinforce each other: researchers can pursue more questions, learn more quickly from negative results, and move promising ideas toward reproducible evidence. This case study provides a concrete foundation for exploring that future and for accelerating investigation and discovery through collaboration among human researchers and their evolving teams of AI specialists.

\end{multicols}
\clearpage
\appendixtitles{yes}
\appendixstart
\appendix
\section{Research workflow: a sequential account}
\label{app:workflow_evidence}

This appendix follows the study from the human research brief through numerical work, critique, and manuscript preparation. It is based on the retrospective summary prepared by the Orchestrator and supplied by us; it is not a verbatim conversation transcript. The stages organize the reported work into a readable sequence, while recognizing that some specialist tasks overlapped. We worked collaboratively in directing and reviewing the research; the role names identify the agents' assigned responsibilities.

\subsection{Setting the research brief and assigning responsibilities}
\label{app:workflow_brief}

We supplied the scientific question, the synthetic spectral database, prior notebooks, and the existing research framework. The project combined two purposes: carrying out an abundance-conditioned spectral-modeling case study and documenting how we could coordinate a team of AI specialists. The manuscript's central contribution was the research workflow, with the astrophysics providing substantive tasks and constraints.

The Orchestrator translated the brief into proposed milestones. We approved or adjusted their scope before specialist work was assigned. The plan connected data preparation, Python/Jupyter execution, scientific checks, figure production, and manuscript assembly. Authorization to assess a model was treated separately from authorization to train it. We also requested visible notebook execution so that numerical work could be inspected. We provided the research direction and remained responsible for scientific decisions and the final manuscript.

\subsection{Inspecting the data and running the training workflow}
\label{app:workflow_data}

The Curator and Python/Jupyter Specialist worked with the database and notebooks we supplied to prepare the numerical study. Their tasks connected the stored stellar parameters and abundances to the nine-input design, organized the spectra and data partitions, and maintained the preprocessing information needed for training and assessment. The nine stellar labels were already part of the scientific dataset; the bots' contribution was to identify and use the relevant fields correctly and turn the supplied material into an executable modeling workflow.

The Python/Jupyter Specialist implemented and ran the training calculations. During training, optimization adjusted the network's weights and biases by comparing its generated spectra with the reference spectra. These learned quantities are the model parameters determined through training, distinct from the stellar labels supplied as inputs. Python also supplied the layer dimensions and parameter-count information used by the Artist to explain the architecture. For assessment, the saved model weights had to be paired with the corresponding training scalers and test selection. The Orchestrator coordinated these numerical outputs with the Critic's checks and the Artist's figures, allowing the specialists to build on one another's work.

\subsection{Detecting and correcting the scaler mismatch}
\label{app:workflow_scalers}
\label{sec:ai_verify}

The scaler episode illustrates how the bots identified a problem in their own numerical workflow and corrected it. An assessment of the \texttt{CondGen} baseline produced a test MAE of approximately $0.024$. The Python/Jupyter Specialist and the Critic checked the evaluation preprocessing against the training setup and identified a mismatch between the scaler files being used. Python corrected the evaluation by using the training-matched scalers; the resulting baseline assessment gave an MAE of $0.003681$. The Orchestrator communicated the diagnosis and corrected result to us, and we approved using the training-matched assessment in the manuscript.

The technical detection and correction therefore arose within the bot team: a numerical result prompted criticism, the specialists checked its cause, and the evaluation procedure was revised. Our role was to review the explanation and approve how the result was reported. The mismatched scalers were stored in \texttt{float64}, whereas the training-matched scalers used \texttt{float32}. Because both scaler identity and numerical precision differed, the approximately $6.5$-fold change in MAE demonstrates the importance of matching preprocessing to training; it does not show that lower numerical precision is intrinsically more accurate.

\subsection{Reviewing abundance diagnostics}
\label{app:workflow_abundances}

The numerical and figure workflow produced abundance sweeps to examine the \texttt{CondGen} elemental response. The Critic reviewed the interpretations of the selected Sc, Ti, Cr, Fe, and Mg diagnostics. In the cool-star Mg sweep, the generated mean flux rose between approximately $+0.67$ and $+1.37$\,dex after decreasing at lower abundances. The bots flagged this non-monotonic response for investigation. At the held atmosphere, the database is reported to contain an exact \texttt{SYNSPEC} spectrum only for the reference Mg abundance of $+0.67$\,dex (id~113144). The additional matching synthesis calculations were not performed, so Figure~\ref{fig:assess_abundance_sweep_Mg_baseline} presents the generated response alone and does not establish a physical-model discrepancy.

The bot team identified the reversal in its own diagnostic output, and the Critic challenged presenting the sweep as a clean demonstration of physical fidelity. The Orchestrator communicated the finding to us, and we approved retaining the plot with an explicit distinction between the observed behavior of the network and the reference comparison needed to interpret it. It then coordinated this decision with the manuscript and figure workflow. This episode documents the bots' detection of a potential problem and their contribution to scientific scrutiny.

The finding arose from the agents' analysis; our role was to review and approve its reporting. Neither the physical cause of the reversal nor its departure from a matched synthesis has been established. Extending the \texttt{SYNSPEC} sweep at the same atmosphere, other abundances, spectral resolution, broadening, and normalization would permit that comparison. The current diagnostic remains a qualitative example at one atmospheric anchor.

\subsection{Comparing alternative architectures}
\label{app:workflow_architectures}

The Python workflow assessed four alternatives to the dense baseline: a wider MLP, wavelength attention, multi-scale Conv1D, and FiLM residual. The Orchestrator reported that none displaced the baseline in the short pilot. We authorized presentation of the comparison with explicit limits and retained the baseline. Section~\ref{sec:arch_comparison} reports the scores for the $256$-spectrum subset. This was an exploratory comparison used in model selection; it did not constitute an independent final validation of the full parameter space.

The team therefore used the comparison to support a project decision. Complete per-pilot training budgets, stopping criteria, and the exact subset-selection record are unavailable in the supplied evidence, so the result does not establish a general architectural ranking. The retrospective account also reports exclusion of failure panels from the figure set. The poor scores and reported FiLM instability remain disclosed, but the missing panels limit inspection of that failure. These limits belong to the evidence for the decision, rather than being evidence that the baseline is universally preferable.

\subsection{Coordinating figures, references, and manuscript review}
\label{app:workflow_manuscript}
\label{sec:ai_pipeline}

The Orchestrator coordinated a manuscript group involving the Editor, Literature Scout, Critic, Artist, and Expert Spectroscopy Reviewer. A separate exchange connected Python and the Artist for numerical and parameter information. Python supplied the values; the Artist used those values to prepare diagrams and diagnostic panels; the Editor inserted the resulting assets into the manuscript. Shared file names and artifact paths connected these steps.

The account describes corrections to overlapping labels and figure borders, as well as a bibliography correction to the LRPayne publication year. The Critic challenged particular claims or displays, while the Expert Reviewer had a distinct manuscript-level role when assigned. We directed the presentation toward the contribution of the AI research team and approved the scientific and editorial choices. Most exchanges were short assignments, acknowledgments, file paths, review labels, or completion reports; the evidence does not describe extended debates between all agents.

The Editor checked figure inclusion and compilation, and the Auditor reported packaging and figure-regeneration checks. These reports support the account of coordinated production but do not replace inspection of the underlying outputs.

\subsection{Applying the nine-input design throughout the workflow}
\label{app:workflow_status}

We specified nine inputs from the outset: effective temperature, surface gravity, projected rotational velocity, microturbulence, and the abundances of Mg, Sc, Ti, Cr, and Fe. Resolving power was fixed at $R=50\,000$ as a property of the spectral database and was therefore kept outside the input vector. The bots worked within this original scientific specification when preparing the numerical implementation and describing the model.

The Python/Jupyter Specialist supplied the architecture information, including the reported total of $274\,182\,496$ model parameters. The Artist used this information to construct the nine-input architecture diagram, and the Editor incorporated the model description into the manuscript. The Orchestrator coordinated these handoffs so that the numerical work, figure, and written explanation followed the same design. This stage shows how the bots translated a shared research specification into connected computational and communication tasks.

Taken together, these stages show the bots preparing and running numerical work, exchanging results, identifying problems, and revising calculations and their presentation. We provided the scientific direction, reviewed consequential decisions, and remained responsible for the manuscript. The appendix documents selected episodes of this collaborative process from the Orchestrator's retrospective account; complete execution records and measurements of human effort, runtime, and cost would support a more systematic evaluation of its reproducibility and efficiency.

\clearpage
\begin{adjustwidth}{-\extralength}{0cm}
\reftitle{References}
\externalbibliography{yes}
\bibliography{project2_core-Version2}
\end{adjustwidth}

\end{document}